# Elasto-visco-plastic flows in benchmark geometries:
# II. Flow around a Confined Cylinder

*Milad Mousavi, Yannis Dimakopoulos, John Tsamopoulos\**

Laboratory of Fluid Mechanics and Rheology, Dep. of Chemical Engineering

University of Patras, Greece, 26504

## *Abstract*

We examine computationally the two-dimensional flow of elastoviscoplastic (EVP) fluids around a cylinder symmetrically placed between two plates parallel to its axis. The Saramito-Herschel-Bulkley fluid model is solved via the finite-volume method using the OpenFOAM software. As in viscoplastic materials, unyielded regions arise around the plane of symmetry well ahead or behind the cylinder, as two small islands located above and below the cylinder and as polar caps at the two stagnation points on the cylinder. Most interestingly, under certain conditions, an elongated yielded area around the midplane is predicted downstream of the cylinder, sandwiched between two unyielded areas. This surprising result appears, for example, with Carbopol 0.1% when considering a blockage ratio of 0.5 (the ratio of the cylinder's diameter to the channel's width) and above a critical elastic modulus ($G$>30 $Pa$). An approximate semi-analytical solution using the same model, in the region mentioned above reveals that it is caused by the intense variation of the stress magnitude there, which may approach the yield stress asymptotically either from above or below, depending on material elasticity. The drag coefficient on the cylinder increases with yield stress and blockage ratio but decreases with material elasticity. The unyielded regions expand as the yield stress increases. They also expand when material elasticity increases because this allows the material to elastically deform more before yielding. Behind the cylinder, the so-called "negative wake" appears which becomes more intense as elasticity increases. Furthermore, by decreasing the elastic modulus or increasing the yield stress beyond a critical value, the yield surface may exhibit damped oscillations, or irregular shapes even without a plane of symmetry, all under creeping flow conditions. Both properties generate these patterns mainly behind the cylinder, because they increase the elastic stresses and the curvature of the streamlines triggering a purely elastic instability.

# 1 Introduction

Flow of yield stress materials arises in various applications, including geophysics [1], and biology [2]. For industrial processes, such as production, storage, and transportation of these materials, understanding their rheology is essential [3, 4]. In oil recovery and groundwater treatment processes, a detailed examination of their flow and displacement within porous media can lead to enhancing the oil recovery processes [5].

On the other hand, to identify the origins of difficulties in numerical simulations of viscoelastic fluid flows, researchers have proposed a set of simple geometries to study them [6, 7]. The flow around a cylinder confined between two plates parallel to its axis is one such geometry that has been used to study different flow structures of viscoplastic and viscoelastic materials with different constitutive models. In this study, we focus on this flow arrangement of elastoviscoplastic (EVP) materials, which combine properties from the two major categories of non-Newtonian fluids. EVP materials are characterized by a yielding transition; when the stress magnitude exceeds the yield stress, the material behaves like a viscoelastic fluid (yielded), and below that threshold, the material behaves like a hyperelastic solid (unyielded).

This prototype arrangement generates areas where shear or extension dominate or both flow types are present, and it has been widely used in microfluidic channels since the early 2000s. It is devoid of corner singularities, so it is considered a "smooth flow", although the two flow stagnation points at the cylinder surface need particular attention. The flow around the cylinder is strongly affected by the blockage ratio ($B_R = \frac{2R}{w}$), where $R$ is the cylinder radius and $w$ the distance between the plates. When $B_R > 0.5$, the two gaps between the cylinder and channel walls are sufficiently narrow to induce a primarily shear force acting on the cylinder's surface. Conversely, for lower blockage ratios, fluid is more readily allowed to pass through these gaps rather than directly interact with the cylinder's surface [8]. Then extensional flow dominates upstream or downstream the cylinder and around the symmetry plane.

Several researchers [9-12] have reported a distinct flow phenomenon in viscoelastic and elastoviscoplastic fluids near the downstream stagnation point of an obstacle, commonly referred to as the "negative wake" or "extension wake". Moreover, lateral asymmetry of the flow field can be observed if the Weissenberg number surpasses a critical value in the presence of explicit or implicit shear-thinning [13, 14].

Various studies [13, 15-18] have investigated high Weissenberg flows for this geometry. For example, they have found that enhancement of the material elasticity increases the elastic stress components around the cylinder (mainly in the downstream channel close to the rear stagnation point) and considerably elongates the wake behind the cylinder, increasing it over $300R$ in Boger fluid for $Wi = 54.2$ [15]. Others have shown that the flow will become asymmetric when the Weissenberg number exceeds a critical value, and the fluid will flow preferentially either above or below the cylinder [13].

Investigators have also explored the flow of viscoplastic materials around a confined cylinder at a constant flow rate. As the Bingham number increases, the (inelastic) unyielded regions expand. However, even at exceedingly high Bingham numbers, (inelastic) yielded zones can still be observed around the cylinder, because of the strong shear force between the cylinder and the plates [19]. The (inelastic) yielded region around the cylinder has always remained symmetrical [19, 20]. The asymmetrical shape of the unyielded region around the cylinder has also been observed in numerical simulations involving thixo-elasto-visco-plastic materials [21-23].

Recent experimental studies have shown that yield-stress materials may have smaller unyielded regions than expected or no fore-aft symmetry around a cylinder [24], and, additionally, a negative wake under creeping flow past a sphere [25]. All these experimental observations have led researchers to recognize the need to incorporate elasticity into the constitutive model, alongside viscosity and plasticity, to more accurately capture the behavior of such materials [10]. Several EVP models have been proposed [26-28], but the most often used ones are by Saramito [29, 30]. The first Saramito model is a combination of the Oldroyd-B viscoelastic model with the Bingham viscoplastic model [29]. Two years later, Saramito improved this model by introducing shear thinning and bounded elongational viscosity via the Herschel-Bulkley model instead of the Bingham model [30]. Cheddadi et al. [24] were the first to use the Saramito-Bingham model [29] in the foam flow around an obstacle. Very good agreement between predictions using the Saramito-Herschel-Bulkley model and experimental observations has been reported in the optimized shape cross-slot extensional rheometer (OSCER) [31].

This study extends our previous one on the flow in the 4 to 1 planar contraction [32] using again the Saramito-Herschel-Bulkley (SHB) model. First, the numerical method (OpenFOAM) and our formulation are validated by reproducing results from the literature in the confined cylinder geometry using the Oldroyd-B fluid model. The drag coefficient is used to assess the accuracy of the numerical predictions. Subsequently, the fitted rheological parameters of the Carbopol-based EVP material [33, 34] are employed in the simulations. Their effect on the yielded/unyielded areas around the cylinder provides insight into the importance of elasticity in the overall flows and particularly downstream of the cylinder. Moreover, the effect of the blockage ratio is investigated, and the importance of sufficient downstream channel length is noted. Most interestingly, when $Wi$ exceeds a certain value, the flow may remain transient for a long time or wavy yield surfaces may appear although the Reynolds number is zero. Similarly, instabilities occur when $Bn$ increases, but their intensity is lower.

## 2 Problem formulation

The governing equations in dimensional form under the assumptions of isothermal, incompressible and two-dimensional flow and in the absence of body forces are the mass (eq. (1)) and momentum balances (eq. (2)):

$$\nabla \cdot \boldsymbol{u} = 0 \tag{1}$$

$$\rho \left( \frac{\partial \boldsymbol{u}}{\partial t} + \boldsymbol{u} \cdot \nabla \boldsymbol{u} \right) = -\nabla P + \nabla \cdot \boldsymbol{\tau}' \tag{2}$$

The extra stress, $\boldsymbol{\tau}'$, is divided in the contributions from the solvent, $\boldsymbol{\tau}_s$, and the yield material, $\boldsymbol{\tau}$. The constitutive equation of the latter is given by the thermodynamically consistent SHB model [30]:

$$\boldsymbol{\tau}' = \boldsymbol{\tau}_s + \boldsymbol{\tau} \tag{3}$$

$$\boldsymbol{\tau}_s = \eta_s (\nabla \boldsymbol{u} + (\nabla \boldsymbol{u})^T \tag{4}$$

$$\frac{1}{G} \overset{\nabla}{\boldsymbol{\tau}} + \max\left(0, \frac{\|\boldsymbol{\tau}_d\| - \tau_y}{k \|\boldsymbol{\tau}_d\|^n}\right)^{1/n} \boldsymbol{\tau} = (\nabla \boldsymbol{u} + (\nabla \boldsymbol{u})^T) \tag{5}$$

In eq. (5), the first term in the left-hand-side is the elastic contribution and the second one the viscoplastic contribution to the rate-of-strain tensor, the max-term imposes the von Mises criterion for material yielding and the term $\overset{\nabla}{\boldsymbol{\tau}}$ indicates the upper-convected time derivative.

$$\overset{\nabla}{\boldsymbol{\tau}} = \frac{\partial \boldsymbol{\tau}}{\partial t} + \boldsymbol{u} \cdot \nabla \boldsymbol{\tau} - \boldsymbol{\tau} \cdot \nabla \boldsymbol{u} - (\nabla \boldsymbol{u})^T \cdot \boldsymbol{\tau} \tag{7}$$

Also, the deviatoric stress tensor, its second invariant and its trace, respectively, are defined in cartesian geometry as:

$$\boldsymbol{\tau_d} = \boldsymbol{\tau} - \frac{\text{tr}(\boldsymbol{\tau})}{N}\boldsymbol{I}, \quad \|\boldsymbol{\tau_d}\| = \sqrt{0.5(\boldsymbol{\tau_d}:\boldsymbol{\tau_d})}, \quad \text{tr}(\boldsymbol{\tau}) = \tau_{xx} + \tau_{yy} + \tau_{zz} \tag{8}$$

In the equations above, $N$ is the number of the dimensions of the problem geometry, $\eta_s$ the solvent viscosity, $\rho$ the fluid density, $G$ the elastic modulus, $k$ the consistency index, $\tau_y$ the yield stress, $n$ the power-law exponent, $(\nabla \boldsymbol{u})^T$ the transpose of the velocity gradient tensor, and $\boldsymbol{I}$ the unit tensor. Below yielding, the max term in eq. (5) is zero, and the hyperelastic model for small solid deformations is recovered; when $G$ approaches infinity, the Herschel-Bulkley model is recovered.

Although the simulations are carried out in dimensional form, it is easier to understand and explain the predictions by referring to the following five dimensionless groups:

$$Re = \frac{\rho U R}{\eta_0}, \quad Wi = \frac{\lambda U}{R}, \quad Bn = \frac{\tau_y R}{\eta_0 U}, \quad \beta = \frac{\eta_s}{\eta_0}, \quad \varepsilon_y = \frac{\tau_y}{G}, \tag{9}$$

Where $U$ is the average velocity at the entrance of the channel (see Fig. 1), and the total material viscosity is defined as $\eta_0 = \eta_s + k(\frac{R}{U})^{1-n}$. The relaxation time may be defined [30], as:

$$\lambda = \frac{\eta_p}{G}, \text{ where } \eta_p = k(\frac{R}{U})^{1-n} \tag{10}$$

However, this definition makes the relaxation time shear-rate dependent. Introducing it is not really necessary for the present analysis where elasticity is determined by $G$. The dimensionless numbers defined in eq. (9) are the Reynolds number, $Re$, which is the ratio of inertial to viscous forces; the Weissenberg number, $Wi$, is the product of relaxation time with a characteristic shear rate; the Bingham number, $Bn$, is the ratio of yield stress to the viscous stress; $\beta$ is the ratio of the solvent viscosity to the total viscosity, and the yield strain, $\varepsilon_y$, is the ratio of the yield stress to the elastic modulus. When the solvent viscosity is negligible, as it is here, the yield strain can be approximated by the product of the Weissenberg and the Bingham numbers, $\varepsilon_y \approx Wi * Bn$.

The drag coefficient (per unit depth) is calculated as usual, not only to compare with existing results of viscoelastic fluids validating our procedure, but also to determine it for the EVP materials studied herein:

$$C_D = \frac{F_D}{\eta_0 U} = \frac{1}{\eta_0 U} \int ((-p\boldsymbol{I}+\boldsymbol{\tau'}) \cdot \boldsymbol{n}) \cdot \boldsymbol{e_x} \, dS \rrbracket \tag{11}$$

In Eq. (11), $\boldsymbol{n}$ is the vector normal to the surface in question and $\boldsymbol{e_x}$ the unit vector in the $x$-direction.

## 2.1 Geometry, mesh details and solution method

The geometry and related dimensions are illustrated in Fig. 1. The mesh structure is taken to be similar to that in previous studies, but higher refinement is employed close to the cylinder in comparison to these other studies as indicated in the insets of Fig. 1.

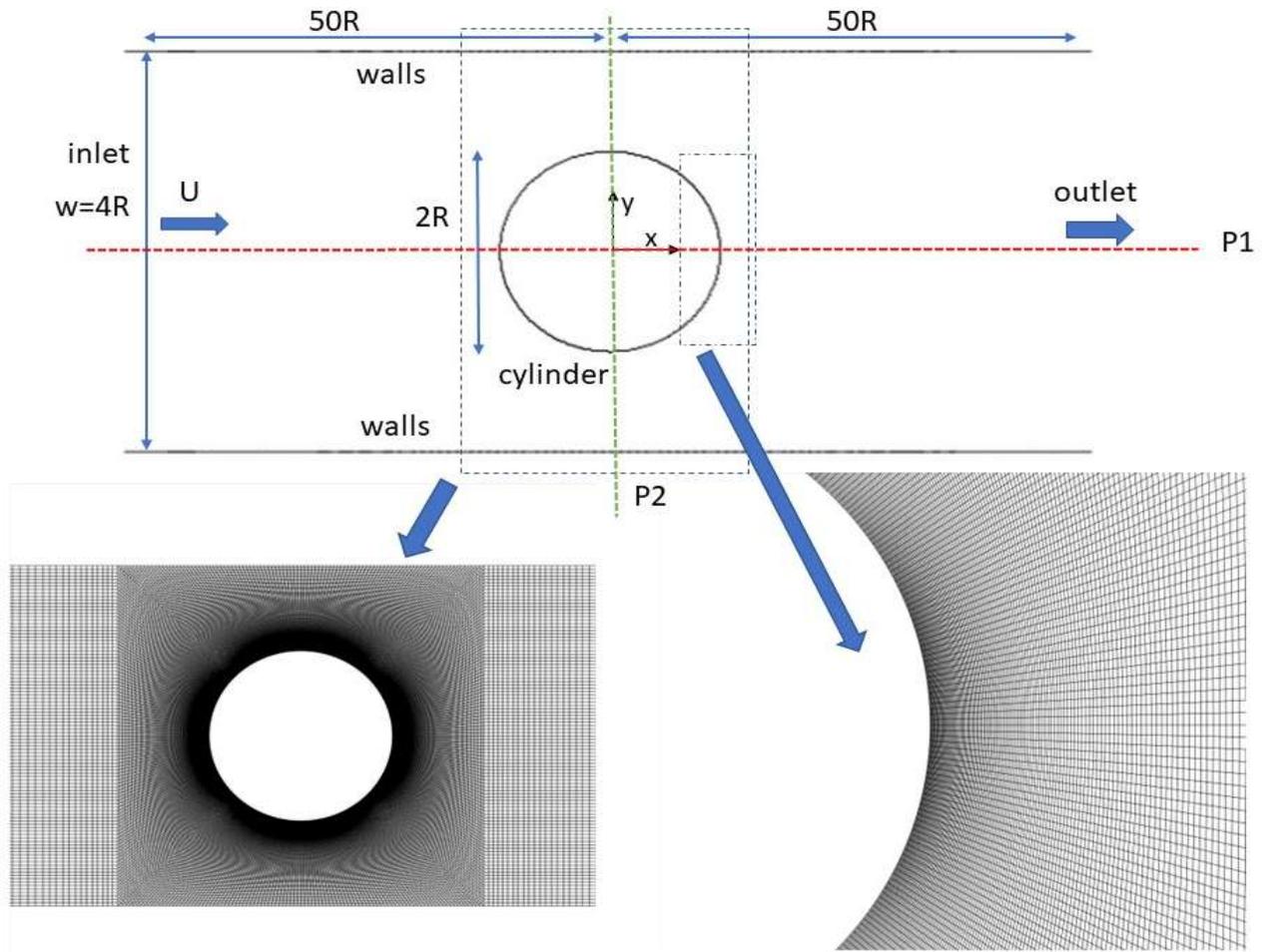

Figure 1. Indicative geometry when the distance between the plates is twice the cylinder diameter. The length of the channel upstream and downstream the cylinder is not drawn in scale. The panel in the bottom left depicts the C3 mesh around the confined cylinder and the panel in the bottom right depicts the mesh refinement around the downstream stagnation point.

The characteristic length for this geometry is the radius of the cylinder, $R$. In Fig. 1, the width of the channel is $4R$ and the blockage ratio is 0.5 ($B_R = \frac{2R}{w}$), while the length of the channel before and after the center of the cylinder is $50R$ (not shown in scale) to minimize, if not eliminate, effects from the inlet and outlet conditions on the flow around the cylinder. The blockage ratio and the channel length have been varied in the parametric study that follows. At the inlet, we apply a uniform velocity profile $U$, zero extra-stress components, and zero-gradient for the pressure. At the outlet, we consider the flow to be fully developed. Therefore, for every variable, we apply a zero-gradient, except for the pressure which is fixed at zero ($p = 0$). On the channel walls and cylinder surface, we apply the no-slip, no-penetration boundary conditions for the velocity and a zero-gradient for the pressure. Also, we use a linear extrapolation of the stress components to the cell boundaries to increase the accuracy of the numerical simulation as suggested in [35]. The symmetry condition in the flow is not imposed on the geometric symmetry plane to allow for possible asymmetries in the flow field.

The governing equations are solved using the same methodology that was briefly described in [32]. For this reason, it will not be repeated here. The only exception is that here the semi-coupled solver, introduced in RheoTool is used. In this approach, the solver is facilitated by the external library (Petsc) for the coupled pressure-velocity fields. Within this framework, all equations associated with the coupled solver are simultaneously solved. Coupled solvers are faster, more stable, and more suitable for transient flow simulations [36]. On the other hand, the stress fields are solved with preconditioned conjugate gradient solver (PBiCG) with the diagonal-based incomplete LU (DILU) preconditioner using iterative (segregated) solver.

## 3 Results and discussion

Our implementation of the OpenFOAM software has been validated by confirming that it can accurately predict the flow of viscoelastic fluids, which has been repeatedly studied; see Appendix A. Moreover, its predictions have been compared with those in [24, 37] which examined the Saramito-Bingham model; see Appendix B.

### 3.1 Flow of an elastoviscoplastic fluid around a confined cylinder

In this section, the flow of the EVP material is examined around a confined cylinder. Initially, we compare our predictions for three Carbopol gels that have been rheologically characterized. Subsequently, we investigate the effect of geometric factors and the four SHB parameters on the results and particularly when the elastic modulus is above, or the yield stress is below a critical value and lastly when these critical values are surpassed.

Tables 1 & 2 present the rheological parameters measured for these three EVP fluids and the corresponding non-dimensional numbers arising in the flow around a confined cylinder. Here, to achieve microfluidic conditions, we have chosen as a characteristic shear rate $U/R = 2s^{-1}$, which implies that for $R = 0.01\ m$, $U = 0.02\ m/s$. Altering the $U$ and $R$ but keeping this proportionality constant does not affect $Bn, Wi,$ and $\beta$. Only the Reynolds number changes in response to such modifications. Notably, the Reynolds number is 0.16 for the Lopez et al. experiment [33] and 0.08 for the Pourzahedi et al. experiment [34]. To expedite the calculations, we set it to zero hereafter, assuming creeping flow. In this study, we introduce a solvent viscosity of $\eta_s = 0.01\ Pa\ s$ - the value for water. As stated in [32], a nonzero $\eta_s$ is required by OpenFOAM. This value of $\eta_s$ results in a fairly small viscosity ratio $\beta = O(10^{-3})$, which is in line with neglecting the solvent viscosity from the rheological characterization of these fluids [33, 34, 38]. It is important to note that all the rheological parameters listed in Table 1 were obtained exclusively from shear rheology tests.

| Rheological Parameters | $n$ | $\tau_y\ (Pa)$ | $k\ (Pa s^n)$ | $G\ (Pa)$ |
|---|---|---|---|---|
| Lopez et al. [33] (0.09%) Fluid 1 | 0.45 | 3.08 | 1.75 | 20.55 |
| Lopez et al. [33] (0.1%) Fluid 2 | 0.46 | 4.71 | 1.81 | 40.42 |
| Pourzahedi et al. [34] (0.1%) | 0.42 | 4.61 | 3.4 | 41.8 |

Table 1- Rheological parameters of Carbopol 0.09% and 0.1% concentrations taken from [33, 34, 38]. Hereafter the first two fluids will be called Fluid 1 and Fluid 2.

| Source of Experimental Data | $Re = \dfrac{\rho U R}{\eta_0}$ | $Wi = \dfrac{\eta_p U}{GR}$ | $Bn = \dfrac{\tau_y R}{\eta_0 U}$ | $\beta = \dfrac{\eta_s}{\eta_0}$ | $\varepsilon_y = \dfrac{\tau_y}{G}$ |
|---|---|---|---|---|---|
| Lopez et al. [33] (0.09%) Fluid 1 | 0.165 | 0.1163 | 1.277 | 0.00829 | 0.1498 |
| Lopez et al. [33] (0.1%) Fluid 2 | 0.159 | 0.0615 | 1.876 | 0.00796 | 0.1165 |
| Pourzahedi et al. [34] (0.1%) | 0.087 | 0.1088 | 1.008 | 0.00437 | 0.1102 |

Table 2- Dimensionless numbers for flow around the confined cylinder for Carbopol 0.09% and 0.1% concentrations taken from [33, 34]. Hereafter Reynolds will be assumed to be zero.

We conducted a mesh convergence study focused on the flow around the confined cylinder for EVP materials. Six mesh configurations were employed for this geometry, and their details are outlined in Table 3. Each subsequent mesh was generated by primarily locally refining the previous one, so we avoided increasing the computational cost excessively. For all comparisons we use the fluid 2 properties obtained via non-linear regression on the data reported by Lopez et al. [33].

| Mesh | Number of cells | Number of points | Number of cells on the surface of the cylinder | Number of y-cells at any cross section away from the cylinder |
|---|---|---|---|---|
| D1 | 40000 | 81000 | 320 | 80 |
| D2 | 71000 | 199962 | 400 | 100 |
| D3 | 149800 | 301260 | 560 | 140 |
| D4 | 390000 | 782700 | 800 | 200 |
| D5 | 552000 | 1107960 | 1600 | 400 |
| D6 | 815000 | 1634860 | 2000 | 500 |

Table 3- Characteristics of the meshes used in the flow of an EVP fluid around the confined cylinder. The first three meshes were used in cases when our simulations converged to a steady state; the latter three meshes were needed when the flow remained time dependent.

In Fig. 2, the profiles of the dimensionless velocity ($\dfrac{u_x}{U}$) and normal stress component ($\dfrac{\tau_{xx}}{\tau_y}$) around the top surface of the cylinder are illustrated. Note that the stress has been normalized by the yield stress in this figure and hereafter.

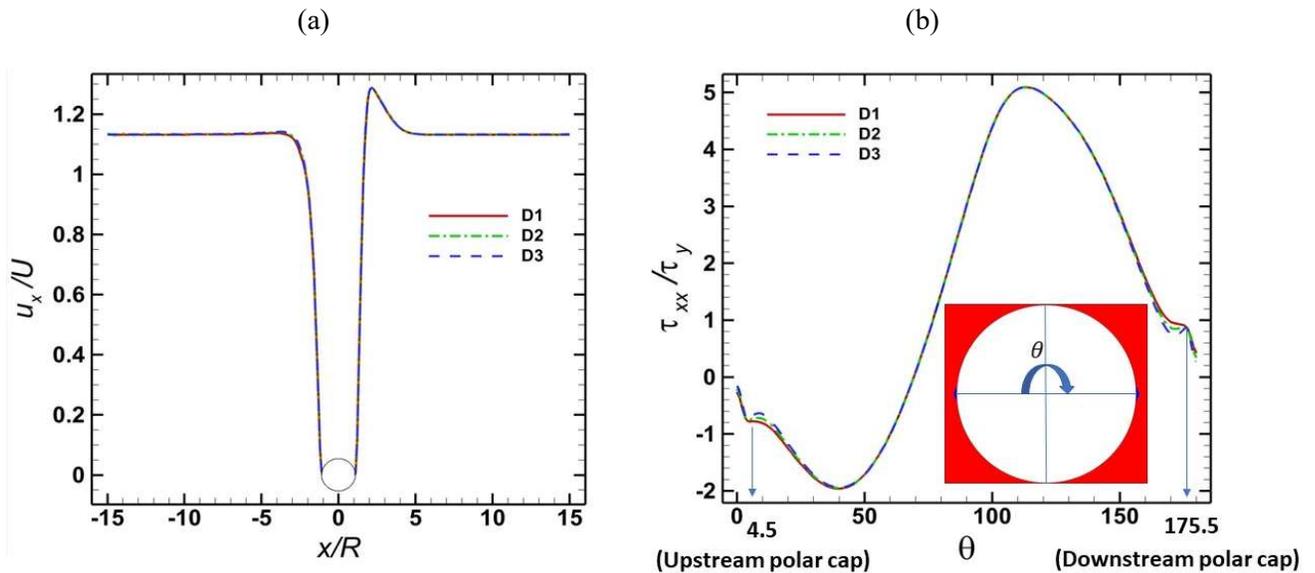

Figure 2. Predicted (a) dimensionless velocity ($u_x/U$) along the plane of symmetry, and (b) dimensionless normal stress ($\tau_{xx}/\tau_y$) around the top surface of the cylinder. The two blue arrows in (b) indicate the points of inception of the two polar caps. The rheological parameters are those of fluid 2, in table 2.

Fig. 2(a) shows the dimensionless velocity along the plane of symmetry. Far upstream and downstream the velocity reaches the same value, which, of course, is higher than the uniform velocity at the entrance, $U$, and becomes zero at the cylinder surface, as it should. A distinct overshoot is observed downstream from the cylinder, which is the sign of the so-called negative wake arising there. The term "negative" becomes clear considering a reference frame in which the far field fluid velocity is zero, then the fluid velocity around the maximum will be in the opposite direction from the cylinder. The negative wake has been previously reported in both viscoelastic and elastoviscoplastic studies [10, 11] and has been attributed to a combination of shear-thinning and axial and shear stresses [9, 12]. All variations of velocity are accurately captured by the first three meshes in Table 3. These meshes will be used until $G$ and $\tau_y$ are beyond certain values and the flow remains oscillatory or erratic.

Fig. 2(b) depicts the dimensionless normal stress component ($\tau_{xx}$) around the top half-surface of the cylinder. It is observed that the material experiences compression at the front of the cylinder and elongation at its back side. It is important to note that in viscoplastic materials and under creeping flow, $\tau_{xx}$ is symmetric with respect to the plane at $\theta = 0^o$ [20]. However, the presence of elasticity in the SHB model breaks this symmetry and $\tau_{xx}$ starts becoming positive earlier, at an angle $\theta \sim 70°$. Furthermore, two small regions of unyielded material, known as polar caps, are visible at the front and back of the cylinder, as with the simpler viscoplastic models. This unyielded region can be identified in Fig. 2(b) by the strictly linear variation of $\tau_{xx}$ observed in the intervals $0^o \leq \theta \leq 4.5°$ and $175.5^o \leq \theta \leq 180°$. This axial normal stress remains nonzero at the two poles. All stress variations are accurately captured by the three meshes in Table 3 as well. Hence, in the subsequent sections, mesh D2 is used for cases leading to a steady solution. When the flow remains transient or becomes periodic or even erratic, the even finer meshes D5 and D6 were examined, to capture flow field complexities more accurately. A transient mesh convergence study for such flows is provided in the Appendix D. Such transient cases are presented in sections 3.5 & 3.6.

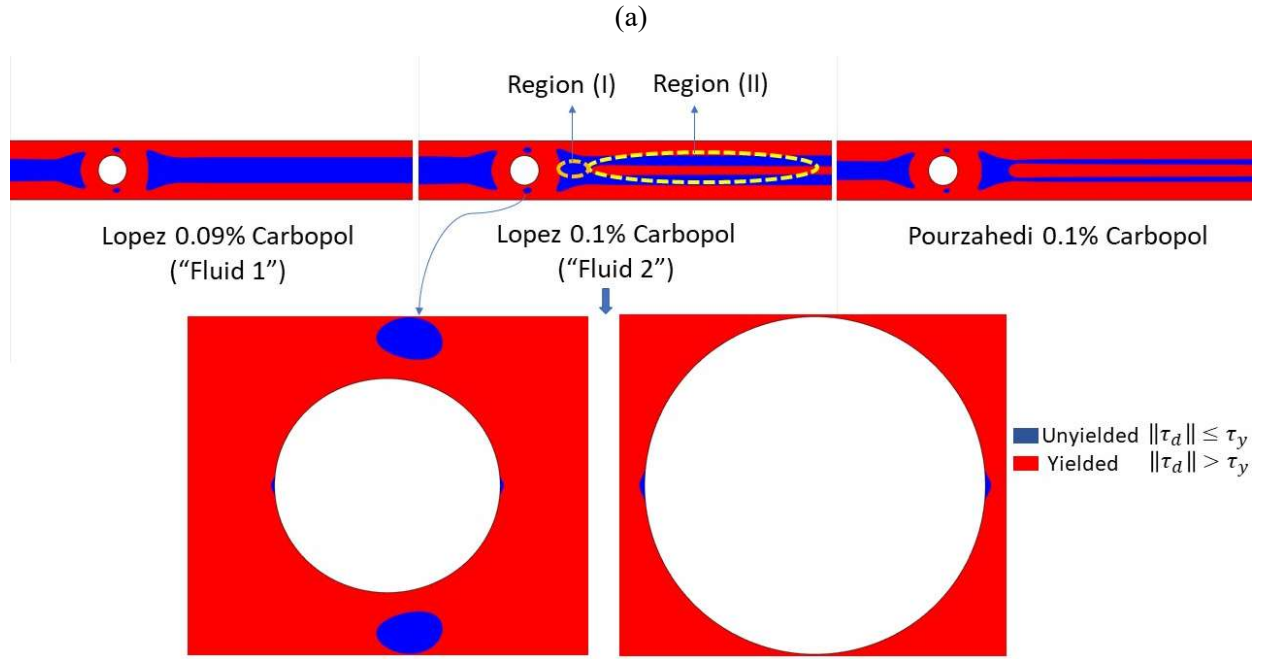

(a)

(b)

Figure 3. a) Predicted yielded (red) and unyielded (blue) material between the cylinders, the range shown is $-7.3 \leq \frac{x}{R} \leq 21.3$, $B_R = 0.5$, for the three sets of rheometric data in Table 1. All subsequent similar figures displaying yielded/unyielded regions have the same axial range. b) Closeups of the area around the cylinder to clarify the islands and the polar gaps for fluid 2, in table 2.

In Fig. 3 (a) the yielded and unyielded areas around the cylinder of the EVP material are shown for the three Carbopol compositions mentioned previously and blockage ratio 0.5. In all cases, the unyielded area ahead of the cylinder is as expected, symmetric and surrounded by yielded material next to the wall, where higher stresses arise. This is qualitatively similar to the corresponding shape in the standard viscoplastic fluids [19, 39]. On the contrary, behind the cylinder, we predict very unexpected shapes for the unyielded area. Whereas the unyielded material for the fluid 1 is as expected and similar to the one for a Herschel-Bulkley fluid, for the other two materials it surrounds yielded material located around the plane of symmetry between the plates. Interestingly, the same unyielded area has been reported before; see fig. 9 in [23], where the entire length of the domain was much larger, $250R$. Possibly because of the smaller blockage ratio in [23], the unyielded area was shorter. The effect of $B_R$ on this unyielded area will be examined in §3.3. We will explain the surprising elongated yielded region that forms behind the cylinder in §3.2. Returning to Fig. 3 (b), we observe two small islands above and below the cylinder and slightly displaced downstream from the plane of symmetry. These result from the presence of a pressure-driven flow between the cylinder and each plate, the transient nature of the flow there (in a Lagrangian frame) and the elastic nature of the material. In the same figure, we may distinguish the very small unyielded zones around the stagnation points at the cylinder surface, the so-called polar caps, as usual [19].

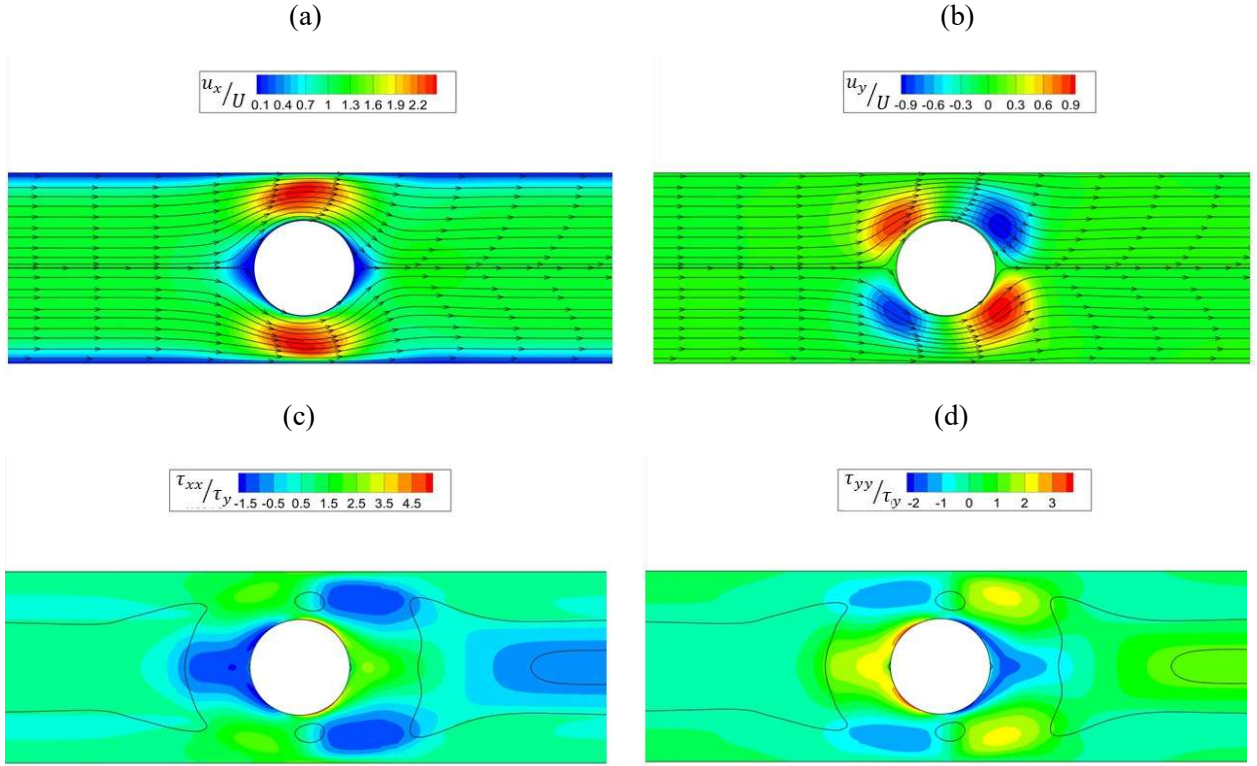

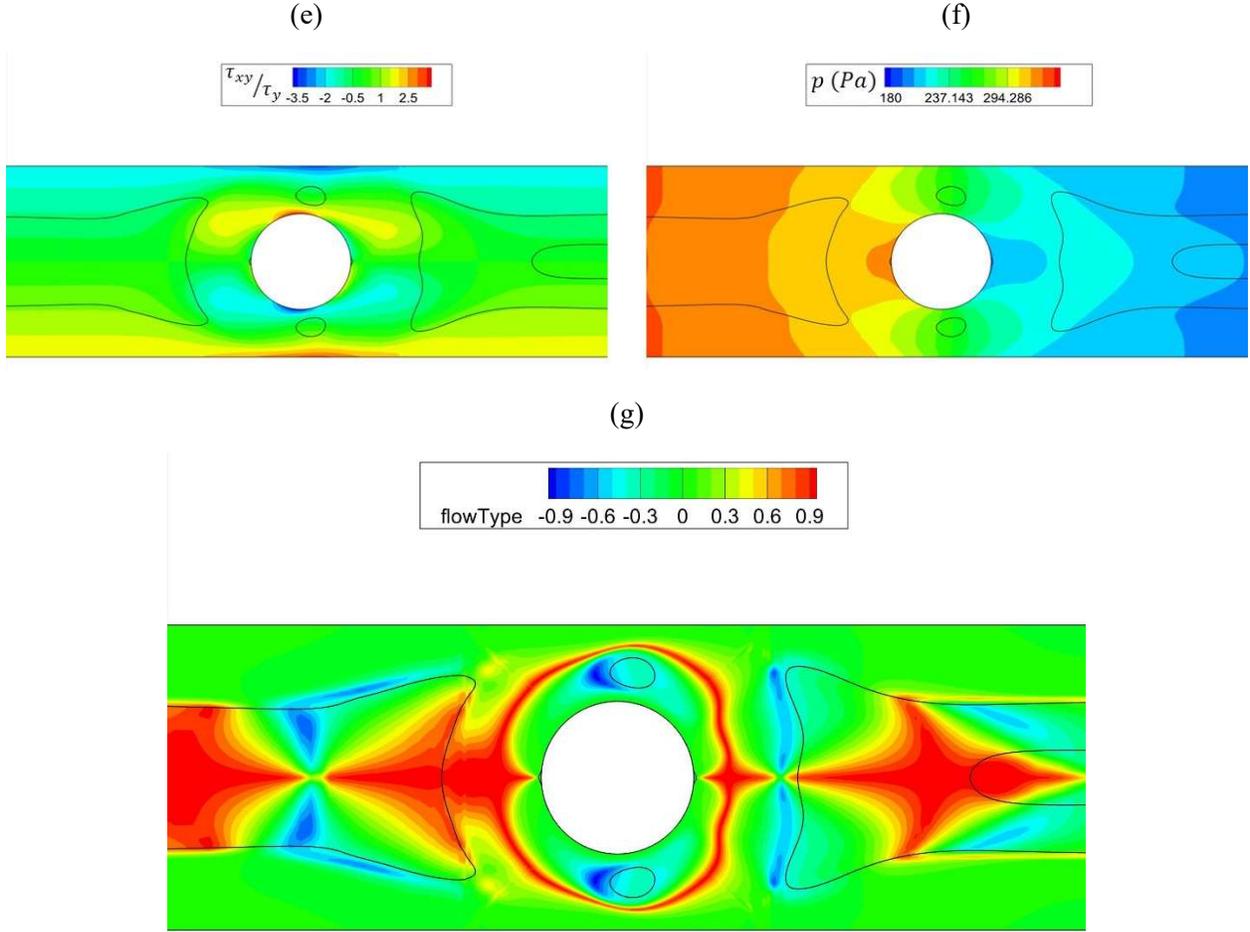

Figure 4. Variation of a) $\frac{u_x}{U}$ b) $\frac{u_y}{U}$ c) $\frac{\tau_{xx}}{\tau_y}$ d) $\frac{\tau_{yy}}{\tau_y}$ e) $\frac{\tau_{xy}}{\tau_y}$ f) $p$, g) flow type parameter ($Q$) around the cylinder ($-6 \leq \frac{x}{R} \leq 6$) for the fluid 2 listed in Table 2. All similar figures presenting stress, velocity, or pressure fields have the same axial range. The locations of the yield surfaces are added as solid black line (see Fig. 3).

Fig. 4 depicts the fields of velocity, stress components, and pressure both upstream and downstream the cylinder. At the upper and lower ends of the cylinder, there are areas of high velocity ($u_x$), because the pathway between the wall and the cylinder's surface is narrowed. Notably, two stagnation points are evident at the front and back of the cylinder, where $u_x$ drops to zero. The secondary velocity component, $u_y$, remains relatively smaller exhibiting also smaller variations, except for four regions near the cylinder at about $\pm 45^o$, where both positive and negative values appear, as expected.

The normal stress component, $\tau_{xx}$ displays negative values at the cylinder's front face due to material planar compression, and, conversely, positive values on its rear surface indicating planar elongation. Further downstream it becomes slightly negative again, a somewhat surprising result indicating elastic rebound. The $\tau_{yy}$ stress component assumes the opposite values to $\tau_{xx}$. Specifically, in the y-direction planar elongation of the material transpires on the cylinder's front surface, whereas compression occurs on the rear surface. The variation of shear component highlights high values near the cylinder's surface and the channel walls. Furthermore, the pressure field illustrates an axial decrease, which well-ahead and after the cylinder becomes almost linear with $x$, as it should.

The type of flow dominating in different parts of the geometry can be determined by the flow-type parameter, $Q$, [40, 41], defined as:

$$Q = \frac{|D|-|\Omega|}{|D|+|\Omega|}, \boldsymbol{D} = \frac{1}{2}(\nabla \boldsymbol{u} + (\nabla \boldsymbol{u})^T), \boldsymbol{\Omega} = \frac{1}{2}(\nabla \boldsymbol{u} - (\nabla \boldsymbol{u})^T), |D| = \sqrt{2(\boldsymbol{D}:\boldsymbol{D})}, \text{ and } |\Omega| = \sqrt{2(\boldsymbol{\Omega}:\boldsymbol{\Omega})}. \quad (11)$$

In the above expressions $\boldsymbol{D}$ is the rate of strain tensor and $\boldsymbol{\Omega}$ is the vorticity tensor. When $Q \sim -1$, the flow is dominated by rotation, when $Q \sim 0$, it is dominated by shear, and when $Q \sim 1$, it is dominated by extension. Fig. 4 (g) shows that extensional flow prevails at the plane of symmetry, both upstream and downstream the cylinder, whereas close to the channel walls, shear flow dominates. Near the cylinder, several rather small regions arise, which are characterized by rotational flow. Two of these areas are manifested above and below the cylinder, where the velocity is maximized. Additionally, two rotational zones appear behind the cylinder, contributing to the formation of a negative wake downstream the cylinder. Moreover, two rotational flows are observable upstream, where the material undergoes compression before traversing the cylinder. On the surface of the cylinder pronounced shearing is predicted, encompassed by a ring-like extension which resembles the experimental observation for viscoelastic fluid [15].

## 3.2 Approximate analysis of the transition zone downstream of the cylinder

Now we will explain the unexpected observations in Fig. 3. EVP materials yield when the von Mises criterion is satisfied. This certainly occurs all along and near the confining plates and the upper and lower side of the cylinder where stronger shear stresses arise and near the front and back stagnation points of the cylinder due to stronger compression and extension normal stresses, respectively. At the plane of geometric symmetry, P1 (Fig. 1), the shear stress is expected to remain small, while the normal stresses should decrease in magnitude as we proceed downstream from the cylinder. Then, one naturally wonders how it is possible to predict a yielded material downstream the cylinder and around the plane of symmetry in the 0.1% Carbopol solutions by both Lopez et al. [33] and Pourzahedi et al. [34]; see Fig. 3 (a). This yielded region extends up to almost the outflow boundary and is "sandwiched" between unyielded zones. This is most unusual and calls for special attention. Two factors contribute to yielding there:

1. Elastic forces: Higher $Wi$ (lower $G$) implies more elastic materials, which deform more prior to yielding and their relaxation time is longer ($G = 20.55 \, Pa$ in the fluid 1). Conversely, materials with smaller $Wi$, are stiffer and, when subjected to stress, they deform less, yield sooner and have shorter relaxation time ($G = 40.42 \, Pa$ in the fluid 2).
2. Transition zones in EVP fluids: As described by Syrakos et al. [42], transition zones appear where yielded materials are trapped within unyielded regions. The EVP constitutive model reveals that the stress magnitude decays exponentially to the yield stress, instead of below it within these transition zones. Hence it formally reaches the yield limit at infinite time or distance.

Based on these observations, we will explain the yielded areas around the midplane for $B_R = 0.5$ by obtaining an approximate analytical solution for the stress field along the $x$-axis in the absence of shear thinning, $n = 1$, (required to make analytical progress). We will base this analytical solution partly on numerical predictions for two Carbopol solutions: the fluid 2 and fluid 1, where a transition zone is and is not predicted, respectively. We will analyze two-dimensional flow only and after a very long time from its initiation, so we may assume that stresses may change only infinitesimally with time ($\frac{\partial \tau}{\partial t} \approx 0$). Then, the constitutive equation simplifies to:

$$\boldsymbol{u} \cdot \nabla \boldsymbol{\tau} - \boldsymbol{\tau} \cdot \nabla \boldsymbol{u} - \nabla \boldsymbol{u}^T \cdot \boldsymbol{\tau} + \max\left(0, \frac{G(\|\boldsymbol{\tau}_d\| - \tau_y)}{k} \frac{1}{\|\boldsymbol{\tau}_d\|}\right) \boldsymbol{\tau} = G(\nabla \boldsymbol{u} + (\nabla \boldsymbol{u})^T) \qquad (12)$$

Also, along the $x$-axis it is expected that $\tau_{xy} \approx 0$, $u_y \approx 0$, and $\frac{\partial u_x}{\partial y} \approx 0$, although planar symmetry was not imposed. It is reasonable to assume that these variables do not vary considerably from these values near the $x$-axis. All these can be seen to hold in Fig. 4 and 5, obtained from the numerical solution, except for very small variations of the normal stresses in the $x$-axis.

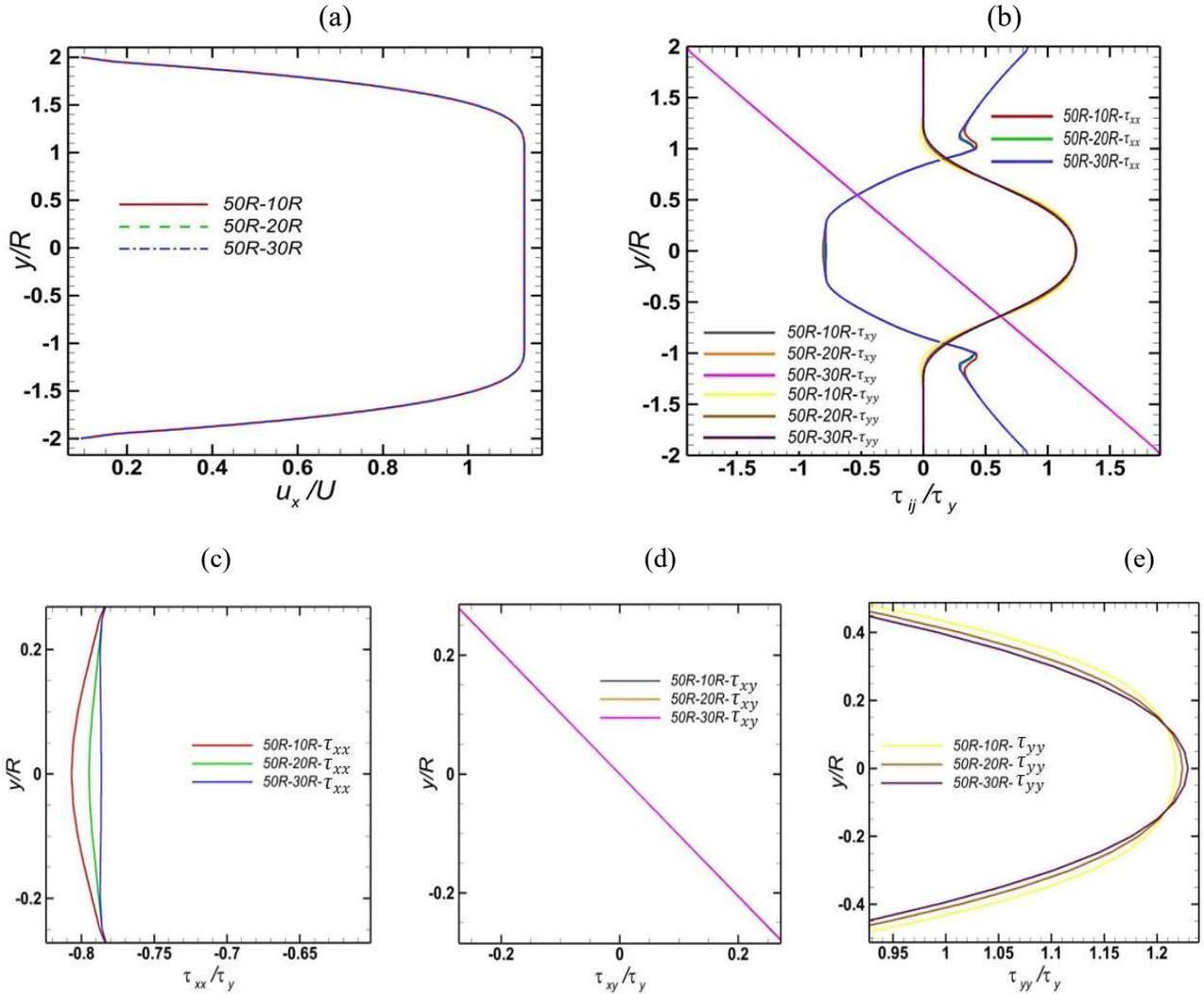

Figure 5. F(a) dimensionless velocity ($\frac{u_x}{U}$), (b) normal and shear stress components over the yield stress (c) $\frac{\tau_{xx}}{\tau_y}$ (d) $\frac{\tau_{xy}}{\tau_y}$ (e) $\frac{\tau_{yy}}{\tau_y}$. The first two figures cover the entire gap between the plates, the latter three figures focus near the mid-plane. In all cases different horizontal locations are examined as indicated with the legends, the first value of which is the location of the exit boundary, and the second value is the $x$-location where the variables were obtained. All numerical data are for the fluid 2 in table 2.

Based on the above, only two components remain from the rate of strain tensor. Using continuity, they are:

$$(\nabla u + (\nabla u)^T)_{xx} = 2\frac{\partial u_x}{\partial x} \tag{13}$$

$$(\nabla u + (\nabla u)^T)_{yy} = -2\frac{\partial u_x}{\partial x} \tag{14}$$

Moreover, observing that $\tau_{xy}$ is the smallest compared to the normal stress components, even away from the $x$-axis, (see Fig. 5) and after making some further simplifications, the magnitude of the deviatoric stress reduces to:

$$\|\tau_d\| \approx \frac{(\tau_{xx} - \tau_{yy})}{2} > 0 \tag{15}$$

We divide our analytical solution into two regions:

- Region I (marked in Fig. 3(a)) corresponds to the unyielded state of the material, where $\max\left(0, \frac{G(\|\tau_d\| - \tau_y)}{k}\frac{1}{\|\tau_d\|}\right) = 0$ in eq. (12). This region exists in both Carbopol solutions.
- Region II (marked in Fig. 3(a)) applies to cases where the transition zone appears, and $\max\left(0, \frac{G(\|\tau_d\| - \tau_y)}{k}\frac{1}{\|\tau_d\|}\right) \neq 0$ in eq. (12). This region exists only in the Carbopol 0.1% solution, whereas Carbopol 0.09% remains unyielded.

**I)** The solution in region I extends from the point where unyielded material is first formed up to either the onset of the transition zone, which occurs for fluid 2 ($2.361 \leq \frac{x}{R} \leq 4.34$), or is extended further downstream until the point where velocity fitting (see below) remains valid ($2.53 \leq \frac{x}{R} \leq 5.5$) for the fluid 1.

The variations along the $x$-axis remain only in the $\tau_{xx}$ & $\tau_{yy}$ components, as shown in Figs. 5(c) and 5(e). Notably, Fig. 5(d) indicates that there is no contribution from $\tau_{xy}$. According to the numerical solution, the primary velocity component $u_x$ even within this unyielded region has a small but important variation (see Fig. 6, below), which can be approximated by a quadratic polynomial, $u_x = Ax^2 + Bx + C$. Then the extension rate component varies linearly $u'_x \equiv \frac{\partial u_x}{\partial x} = 2Ax + B$. These three coefficients are determined by fitting $u_x$ to the numerical predictions and are reported in Table 4.

The two non-trivial components of the constitutive model reduce to:

$$\frac{\partial \tau_{xx}}{\partial x} - \tau_{xx}\left(\frac{2u'_x}{u_x}\right) = \frac{(2Gu'_x)}{u_x} \tag{16}$$

$$\frac{\partial \tau_{yy}}{\partial x} + \tau_{yy}\left(\frac{2u'_x}{u_x}\right) = \frac{-(2Gu'_x)}{u_x} \tag{17}$$

Solving eq. (16) and eq. (17), and applying the numerically obtained values $\tau_{xx_1} \equiv \tau_{xx}(x = x_1)$ and $\tau_{yy_1} \equiv \tau_{yy}(x = x_1)$, at the point where the unyielded material starts to form, $x_1$, yields:

$$\tau_{xx} = G\left(\left(\frac{u_x}{u_{x_1}}\right)^2 - 1\right) + \tau_{xx_1}\left(\frac{u_x}{u_{x_1}}\right)^2 \tag{18}$$

$$\tau_{yy} = G\left(\left(\frac{u_{x_1}}{u_x}\right)^2 - 1\right) + \tau_{yy_1}\left(\frac{u_{x_1}}{u_x}\right)^2 \tag{19}$$

The values $\frac{x_1}{R}, \frac{u_{x_1}}{U}, \frac{\tau_{xx_1}}{\tau_y}, \frac{\tau_{yy_1}}{\tau_y}$ are presented in Table 4. The end of the unyielded area and the beginning of the transition zone, $\frac{x_2}{R}$, is also given in Table 4 for the fluid 2.

| Constants | Fluid 1 | Fluid 2 |
|---|---|---|
| A | 4.612 | 3.167 |
| B | -0.489 | -0.352 |
| C | 0.036 | 0.032 |
| $\frac{x_1}{R}$ | 2.5 | 2.3 |
| $\frac{x_2}{R}$ | - | 4.6 |
| $\frac{u_{x_1}}{U}$ | 1.3 | 1.25 |
| $\frac{\tau_{xx_1}}{\tau_y}$ | 1.319 | 1.196 |
| $\frac{\tau_{yy_1}}{\tau_y}$ | -0.68 | -0.803 |

Table 4- Constants which are obtained from or fitted to the numerical predictions for the fluid 1 and fluid 2.

Figure 6 depicts both the numerical and the analytical solutions. In both panels, the lines vary continuously in the range $1 \leq x/R \leq 10$, while the symbols start where yielded material first appears behind the cylinder and end where the analytical solution ends. First, we may observe the very small variation of $u_x$ after it exhibits the usual maximum indicating the negative wake. The magnitudes of both normal stresses and of $\frac{|\tau_{xx} - \tau_{yy}|}{2}$ increase abruptly, exhibiting also sharp maxima at the same axial position, but earlier than the velocity, and then decrease more smoothly behind the cylinder. These sharp variations and strong interaction with the rest of the fluid in the same cross section make analytical progress in this region impossible.

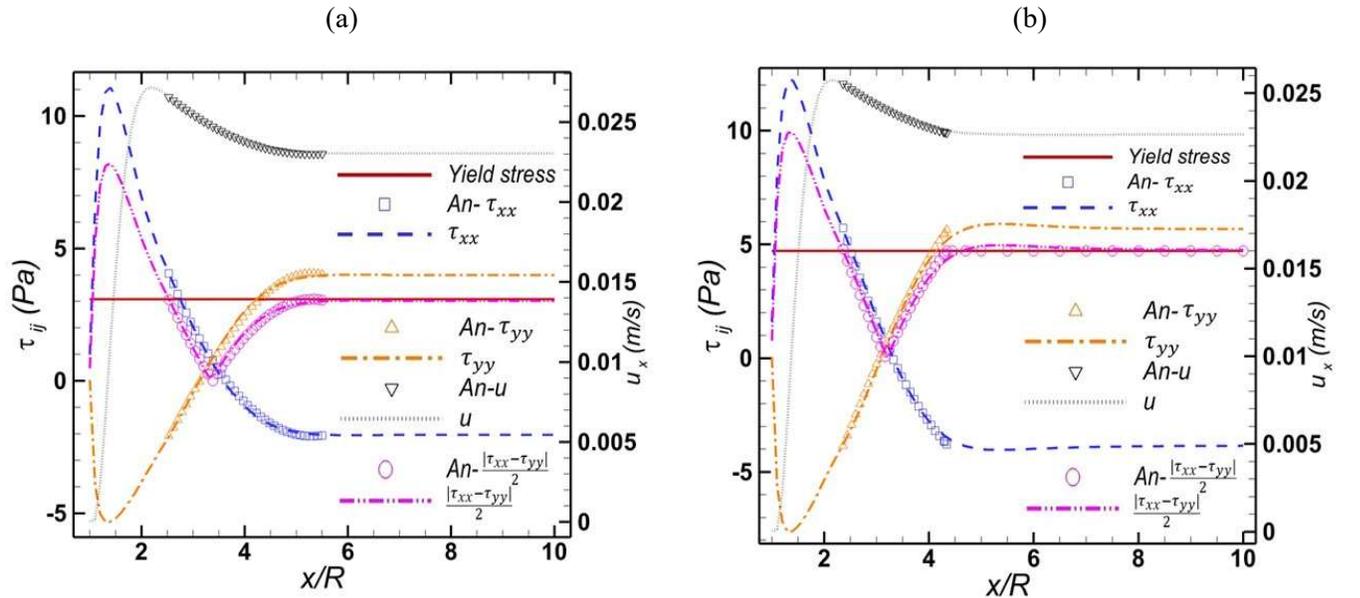

Figure 6. F Comparison of the variation of $\tau_{xx}, \tau_{yy}, \frac{|\tau_{xx}-\tau_{yy}|}{2}, u_x$ at $y = 0$ obtained from either numerical or analytical calculations for a) the more elastic fluid 1 and b) the less elastic fluid 2. Symbols represent analytical solutions and in the corresponding legend they are indicated by "An", while lines represent numerical calculations in both regions (I) and (II).

Examining closer Fig. 6a depicting results for the fluid 1, we observe that it does not exhibit a transition zone since both the line and the symbols for $\frac{|\tau_{xx}-\tau_{yy}|}{2}$ after they fall below the yield stress, approach it again from BELOW asymptotically. This confirms the observation in Fig. 3a. Since this fluid does not exhibit a transition zone but remains unyielded, eqs (20) and (21) hold for as long as $u_x$ can be approximated by a quadratic polynomial. Subsequently, the constant $u_x$, combined with the asymptotes reached by the stress components keep the material unyielded. Throughout, the analytical solution closely follows the numerical predictions.

Fig. 6b shows the corresponding numerical and analytical calculations for the fluid 2. Interestingly, here both the analytical and numerical solution do not approach the yield stress line asymptotically, but cross it, clearly indicating that the material yields at the corresponding location. Hence in this case, eqs (20) and (21) are valid only until $\frac{x}{R} = 4.34$, where the analytical calculation predicts the beginning of the transition zone. This prediction slightly differs from the numerical prediction, which places the beginning of the transition zone at $\frac{x}{R} = 4.63$. Beyond this location, due to the appearance of the transition zone, this analytical solution is no longer valid. To study beyond this point and work in region II one needs to apply the different set of equations described below.

**II**) In this region, ideal plug flow exists (although the material is yielded according to the von Mises criterion), meaning that $u_x$ is approximately constant and $u_y = 0$. Since the velocity gradient is zero, the last two terms in the upper-convected time derivative vanish (refer also to Fig. 5(a)). In this solution, $\max\left(0, \frac{G(\|\tau_d\|-\tau_y)}{k}\frac{1}{\|\tau_d\|}\right) \neq 0$. The same assumptions for the variation of $\tau_{xx}, \tau_{yy}$ & $\tau_{xy}$ along the $x$-axis remain valid as in the first region. Overall, Fig. 5 indicates that the transition zone appears due to slight variations in the normal stress components, although the velocity and shear stress do not vary at all. Accordingly, in the section where the transition zone appears ($4.34 \leq \frac{x}{R} \leq 10$), eq. (12) reduces to:

$$\boldsymbol{u} \cdot \nabla \boldsymbol{\tau} + \frac{G}{k}\left(\|\boldsymbol{\tau}_d\| - \tau_y\right)\frac{1}{\|\boldsymbol{\tau}_d\|}\boldsymbol{\tau} = 0 \tag{20}$$

The magnitude of the deviatoric stress $\|\boldsymbol{\tau}_d\|$ is given by eq. (15) and the $\tau_{xx}$ & $\tau_{yy}$ components are obtained from:

$$\frac{\partial \tau_{xx}}{\partial x} + \frac{\tau_{xx}}{u_x \lambda}\left(1 - \frac{\tau_y}{\left(\frac{\tau_{xx}-\tau_{yy}}{2}\right)}\right) = 0 \tag{21}$$

$$\frac{\partial \tau_{yy}}{\partial x} + \frac{\tau_{yy}}{u_x \lambda}\left(1 - \frac{\tau_y}{\left(\frac{\tau_{xx}-\tau_{yy}}{2}\right)}\right) = 0 \tag{22}$$

Subtracting eq. (21) from eq. (22) yields:

$$\frac{\partial(\tau_{xx}-\tau_{yy})}{\partial x} + \frac{1}{u_x \lambda}\left(\tau_{xx} - \tau_{yy}\right) = \frac{2\tau_y}{u_x \lambda} \tag{23}$$

With solution:

$$\frac{\tau_{xx}-\tau_{yy}}{2} = \tau_y + De^{\frac{-x}{u_x\lambda}} \qquad (24)$$

In eq. (24), the integration constant $D$ must be determined by equating this stress difference with the corresponding one at the end of the first part of the analytical solution. However, upon inserting the value of $\frac{|\tau_{xx}-\tau_{yy}|}{2}$ from the end of solution (I) into eq. (24), it makes apparent that $D = 0$. Consequently, the final form of the solution is $\frac{|\tau_{xx}-\tau_{yy}|}{2} = \tau_y$.

This is a shortcoming of the analytical solution and calls for further examination. The numerical solution for the fluid 2 clearly reveals that $\frac{|\tau_{xx}-\tau_{yy}|}{2}$ undergoes a second albeit smaller than the first maximum. The second maximum is always correlated with the appearance of a transition zone. Hereafter, the second maximum is considered to be the recoil in stress magnitude that occurs after the initial growth and its reduction to zero. As already mentioned, at the end of region I both the analytical and the numerical solution predict that the lines representing this quantity will intersect the yield stress line yielding the material. Moreover, the numerical solution demonstrates that the stress magnitude, after intersecting the horizontal yield stress line, will approach it again asymptotically but from ABOVE. Then, why can't the analytical solution follow suit? The reason for this is that our assumption of $n = 1$ makes the second maximum quite weak and coupling this with our other approximations to derive an analytical solution result in $D = 0$. To confirm this idea, we turn our attention to the effect of $n$ on the general numerical solution to be discussed section 3.4. In Fig. 10b (see section 3.4) we observe that when the effect of shear thinning is neglected, the second maximum in stress magnitude is so small that it cannot be calculated analytically, while when $n$ decreases (indicating more shear thinning materials), the second maximum becomes stronger, and the decay of stress becomes slower and smoother, as also mentioned by Syrakos et al. [42].

In summary, Figs. 6a & 6b show that the analytical and numerical calculations approach the yield stress in two distinct manners:

- For **fluid 1**, the stress magnitude never reaches the yield stress but approaches it from **BELOW**, resulting in no appearance of a transition zone.
- For **fluid 2**, the stress magnitude surpasses the yield stress reaching a second maximum slightly above the yield stress and approaches it again from **ABOVE**, leading to the continuous appearance of the transition zone downstream to a specific distance.

The difference in yield stress between the two Carbopol solutions is rather small, $\tau_y = 3.08\ Pa$ vs. $\tau_y = 4.71\ Pa$ as also seen in Fig. 6 and does not cause the qualitative difference in flow between these two fluids. On the contrary, the difference in elastic modulus $G = 20.55\ Pa$ vs. $G = 40.42\ Pa$ is significant and explicitly enters the analytical solution for the stress components, eqs (18) and (19). Apparently, this property generates the transition zone, when the material becomes less elastic.

## 3.3  Effect of the blockage ratio ($B_R$)

In what follows, unless otherwise stated, and until the end of section 3, the material will be fluid 2 [33], with $B_R = 0.5$, and the location of the outflow boundary at $50R$, while one parameter will be changed in each subsection to examine its effect. All simulations leading to a steady state were terminated at $t = 100$ s. In this section, we analyze the effect of the blockage ratio on the yielded/unyielded zones and the intensity of the negative wake. Additionally, we assess the impact of the confinement on the drag coefficient around the cylinder.

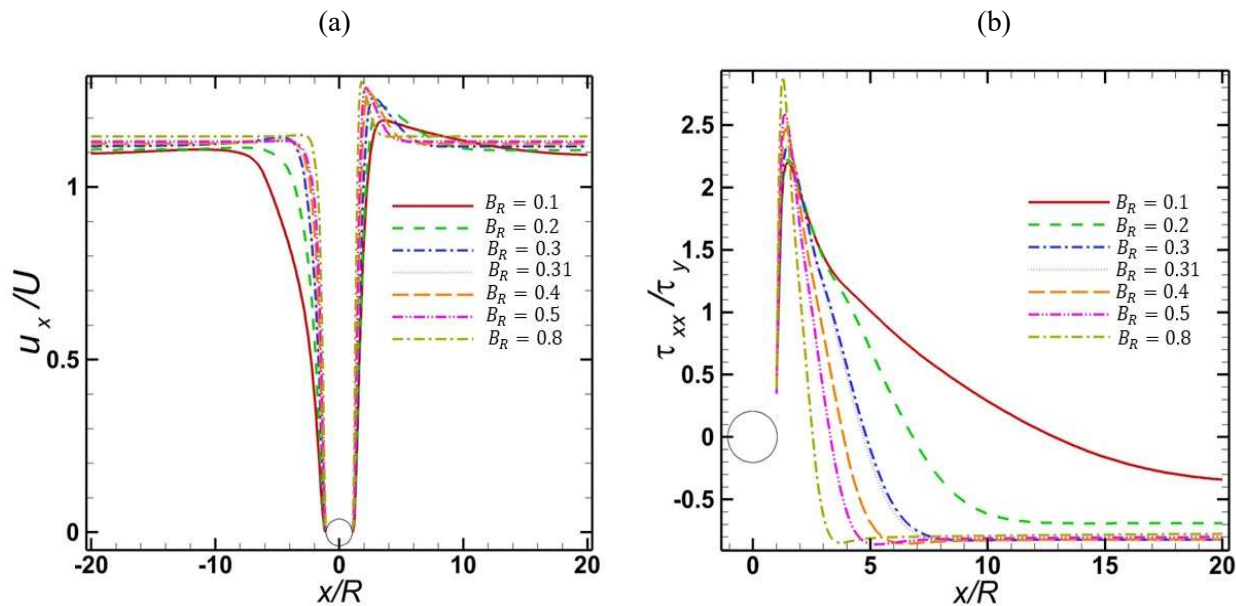

Figure 7. Dimensionless (a) velocity ($\frac{u_x}{U}$) and (b) normal stress component ($\frac{\tau_{xx}}{\tau_y}$) along the $x$-axis and at $y = 0$ for different blockage ratios ($B_R$). The material is fluid 2 in table 2.

Fig. 7(a) illustrates that as the blockage ratio increases, the velocity overshoot behind the cylinder increases and approaches the fully developed value more abruptly. On the contrary, in the lower blockage ratios, the decay of the velocity towards the fully developed value is smoother. This dependence is expected, because the higher $B_R$ generates stronger shear stress in the constriction between the cylinder and the wall and stronger axial normal stress because of a more abrupt expansion when the constriction ends. A higher velocity overshoot indicates a stronger negative wake behind the cylinder. In Fig. 7(b), a similar overshoot is observed for the axial normal stress component. The maximum value of the normal stress component behind the cylinder indicates that the material elongates the most, especially in the case of higher blockage ratios, as just explained. It is evident that this overshoot is more prominent and decays more rapidly for higher blockage ratios. This corresponds to the increase in the stress overshoot when elasticity increases [16]. For $B_R > 0.3$, $\tau_{xx}$ approaches the same negative value far downstream, that is lower than the corresponding asymptote for smaller $B_R$. According to Fig. 7(b), a compression along the symmetry plane occurs after the material undergoes significant elongation.

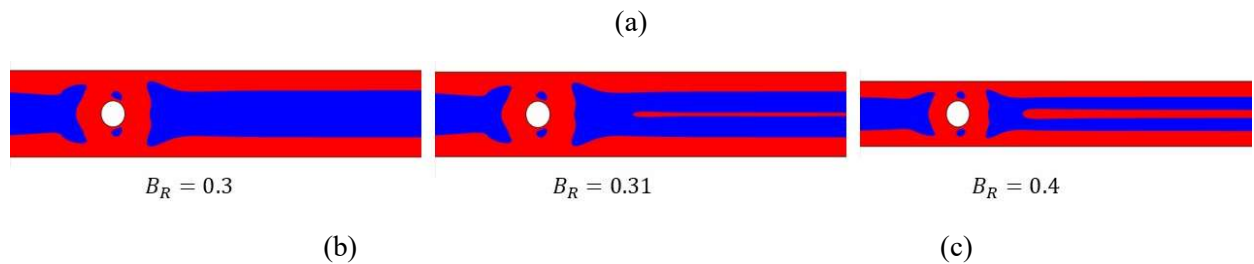

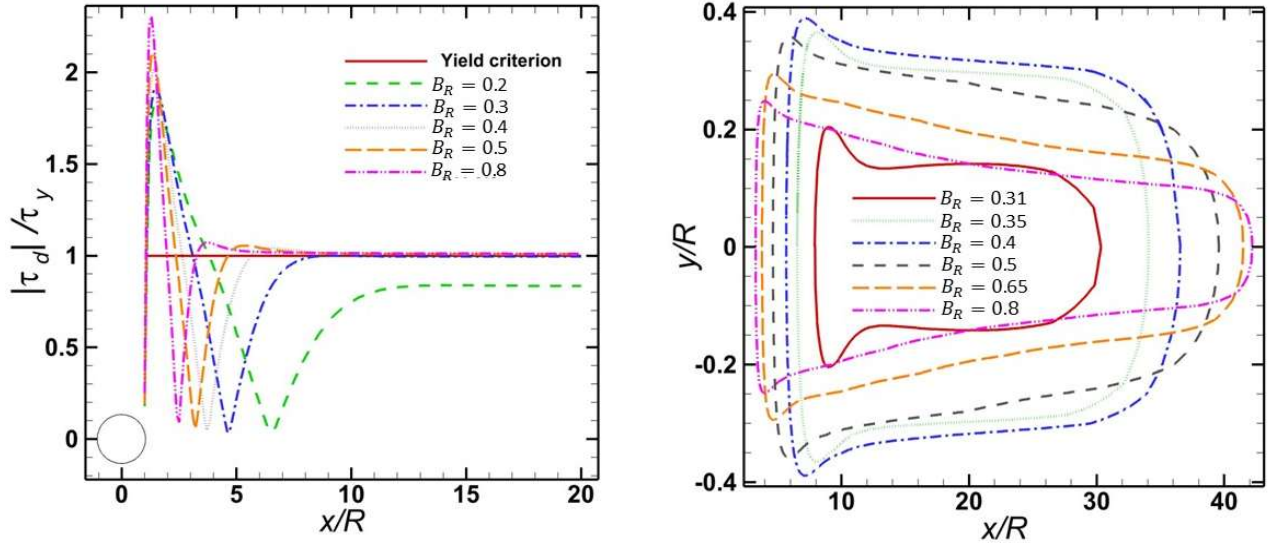

Figure 8. Effect of the blockage ratio on a) the yielded (red) and unyielded (blue) material, and b) dimensionless stress magnitude $\frac{|\tau_d|}{\tau_y}$, and c) the inner yield surface when yielded material arises around the symmetry plane. The material is fluid 2 in table 2.

Fig. 8(a) depicts the yielded/unyielded zones for three blockage ratios. Yielded material around the symmetry plane is observed only when $B_R > 0.3$. Apparently, this is the minimum $B_R$ that induces large enough shear and normal stresses behind the cylinder, which, according to the approximate analysis, can lead to such stress magnitude that exceeds the yield stress before it starts decreasing asymptotically. Fig. 8(b) shows that at $B_R = 0.4, 0.5, and\ 0.8$, a second maximum in stress magnitude emerges, indicating the appearance of the transition zone in the downstream channel. On the contrary, at $B_R = 0.2, and\ 0.3$ the stress magnitude approaches asymptotically from below and remains below the yield stress, indicating that a transition zone will not appear, as seen in Fig. 8a and 8c. Furthermore, Fig. 8(c) shows the inner yield surfaces around the symmetry plane for different blockage ratios. The magnitude of the stress in the domains inside these contours is very close but above the yield limit, so they constitute transition zones. The length of the yielded zones increases monotonically with $B_R$, and in a way that their upstream side approaches the cylinder, and their downstream side approaches the outflow boundary. On the other hand, the thickness of these transition zones increases monotonically until $B_R = 0.4$, but beyond this value it decreases also monotonically. As $B_R$ increases, the second maximum of stress magnitude (see Fig. 8a) becomes stronger, increasing suddenly and then approaching the yield stress downstream smoothly. When $B_R > 0.4$, after its second maximum, the stress magnitude decreases with a steeper slope towards the outflow boundary, which causes the non-monotonic behavior of the thickness of the transition zone afterwards.

Fig. 9 indicates that the drag coefficient on the cylinder increases with increasing $B_R$. As the blockage ratio increases, the distance between the channel wall and the cylinder surface decreases, inducing stronger velocity gradients, which result in higher stresses acting on the surface of the cylinder, and a higher drag coefficient.

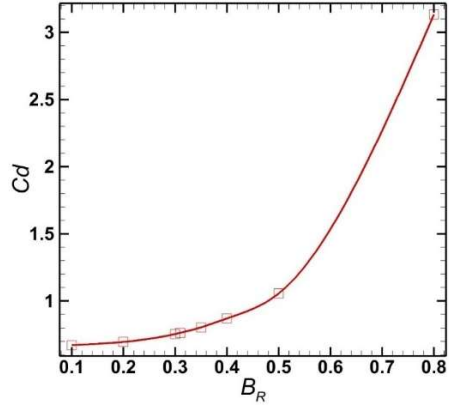

Figure 9. Drag coefficient on the cylinder versus blockage ratio.

## 3.4 Effect of each rheological parameter when a steady solution is reached

First, we will examine the effect of the shear thinning exponent, $n$, on the yielded/unyielded regions. We will restrict the values of the exponent to $n \leq 0.5$. As explained in [13, 43], extension-rate hardening cannot be observed in typical EVP materials, such as Carbopol, emulsions, etc., because of their structure.

(a)

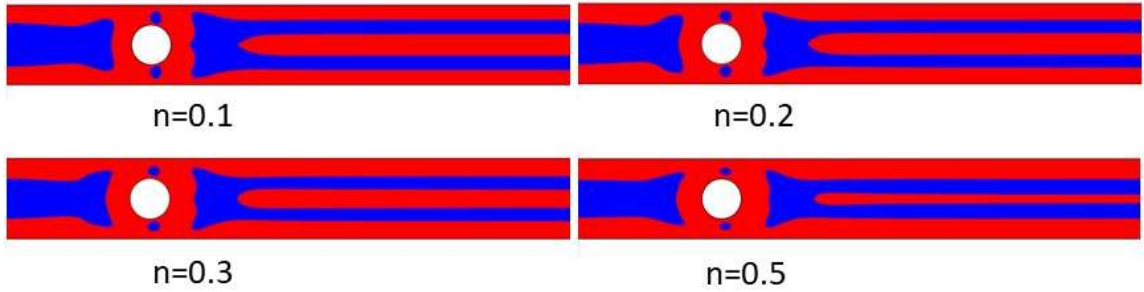

(b) (c)

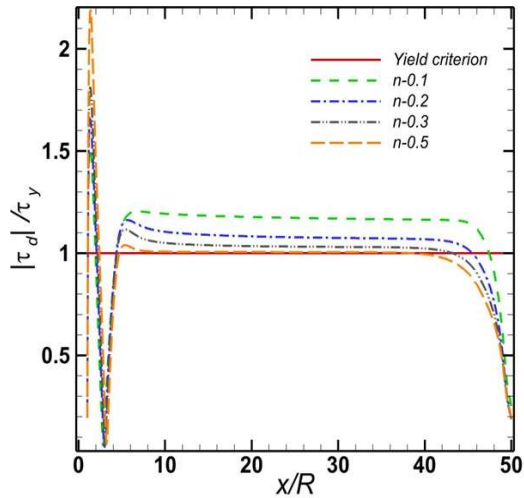
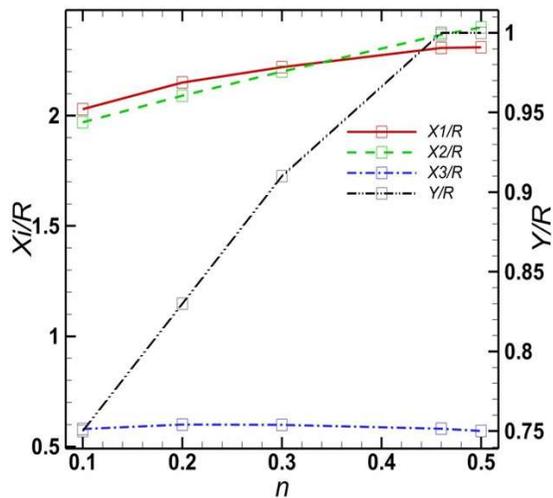

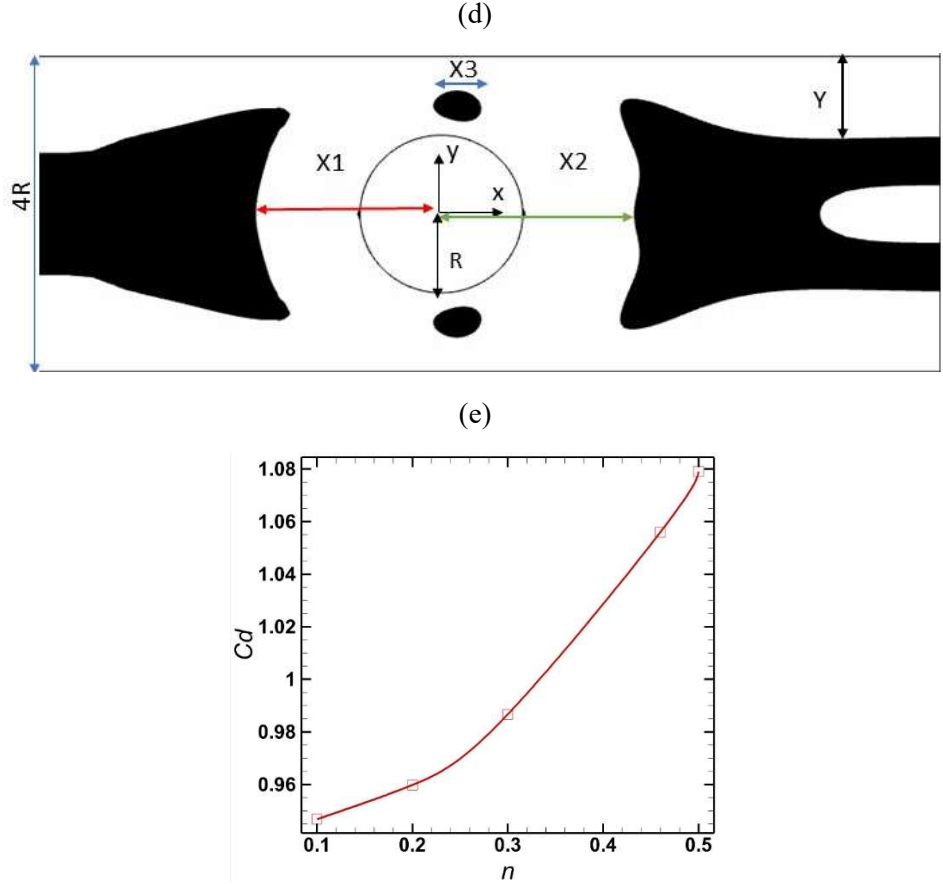

Figure 10. a) Effects of the shear thinning index $n$ on the yielded (red) and unyielded (blue) zones. b) Variation of stress magnitude along $x$-axis at $y=0$ and c) dependence of characteristic lengths on $n$. d) Definition of different horizontal and vertical lengths. e) Drag coefficient on the cylinder versus $n$. The rest of the parameters are those of fluid 2, Table 2.

In Fig. 10 (a) the effect of the shear thinning index $n$ on the yielded/unyielded regions is depicted. According to the definitions of the dimensionless numbers that are given in Table 2, the increase of $n$ decreases the Bingham number and increases the Weissenberg number. As can be seen quantitatively in Table 5, the decrease of $Bn$ in the range of $n$ examined is more important, while $Wi$ increases but remains small.

| $n$ | 0.1 | 0.2 | 0.3 | 0.46 | 0.5 |
|---|---|---|---|---|---|
| $Wi$ | 0.047 | 0.051 | 0.055 | 0.061 | 0.063 |
| $Bn$ | 2.403 | 2.243 | 2.094 | 1.876 | 1.825 |

Table 5. Effect of $n$ on $Bn$ and $Wi$

Hence, the thickness of the yielded area next to the walls should increase because of the decrease of $Bn$. Similarly, the polar caps and islands become somewhat smaller, which resembles the corresponding predictions of viscoplastic materials [19, 44]. Moreover, the thickness of the transition zone around the plane of symmetry decreases, as $n$ increases. This is because increasing the shear thinning of the material decreases the stress components and, hence, the first maximum of the stress magnitude, allowing for a sharper rebound above the yield limit and a higher second maximum, which also takes longer to reapproach $\tau_y$ (see Fig. 10(b)). This idea corroborates with our earlier statement that making the material less shear thinning (higher $n$) results in increased elasticity (higher $Wi$) and reduced stiffness (lower $Bn$). This in turn causes the overall elastic stress components to rise sharply, leading to a higher initial growth of stress magnitude behind the cylinder. This effect extends to a wider part of the cross section, leading to a wider transition zone. It is noteworthy that, according to the analysis in Section 3.2, highly shear thinning materials are more prone to experiencing a stronger second maximum, which consequently leads to the appearance of an increased transition zone (see Fig. 6(b) and 10(b)).

Figure 10 (d) defines certain geometric characteristics of the yield surfaces. $Y$ is the thickness of the yielded film next to the wall downstream of the cylinder where the flow is fully developed. $X_{1,2}$ is the distance of the tip of the main unyielded region upstream/downstream from the center of the cylinder. $X_3$ is the maximum horizontal length of the island above the cylinder. With the increase in $n$, $Y$ increases for the reasons described above. For the same reasons, $X_{1,2}$ increase also as shown in Fig. 10 (c). Meanwhile, the horizontal length of the islands remains nearly constant.

Fig. 10(e) illustrates the increase of the drag coefficient with $n$. This is due to the combined effect of $n$ on $Bn$ and $Wi$ as well as the resulting increase of material viscosity next to the plates, and hence the slightly increased drag coefficient. In any case, the drag coefficient changes only slightly since $n$ varies within the limited range of 0.1 to 0.5.

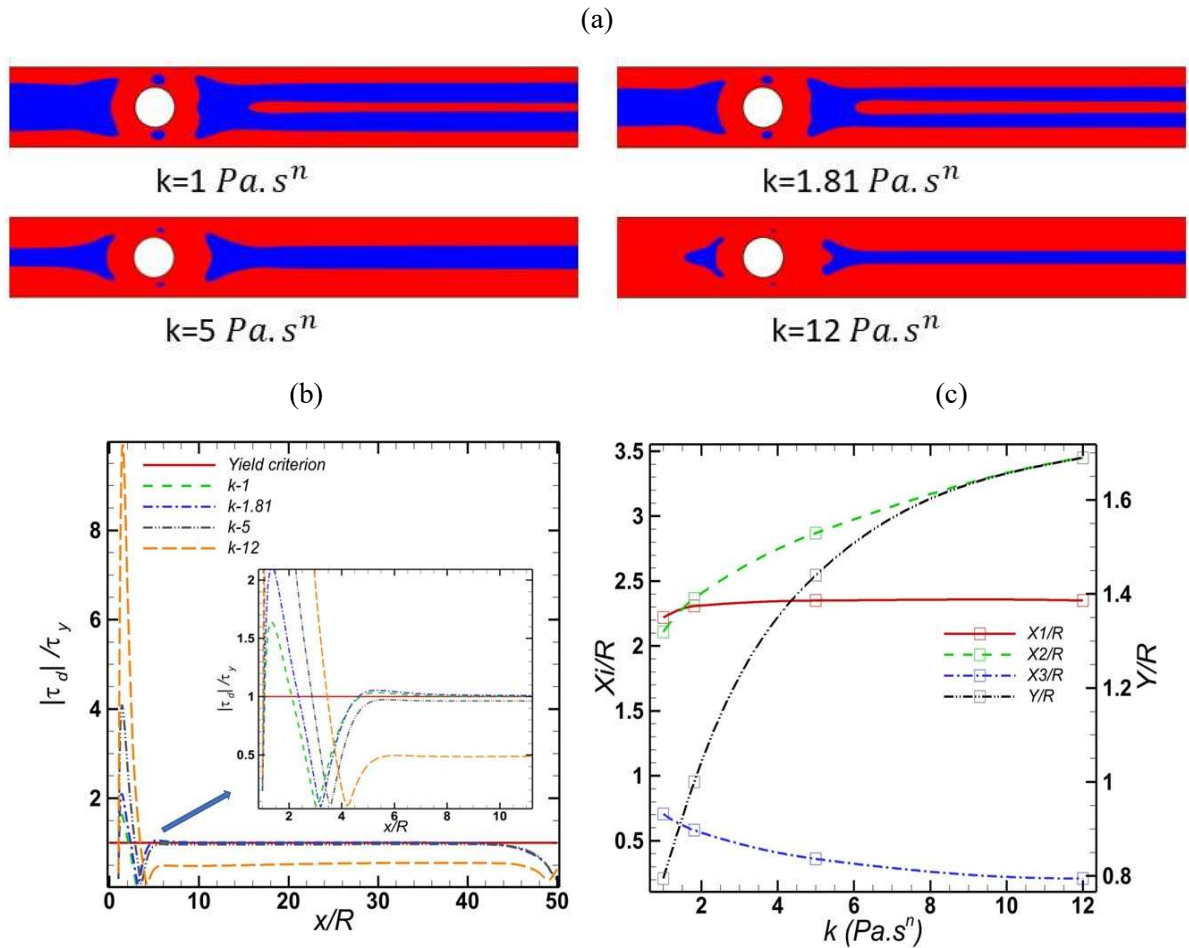

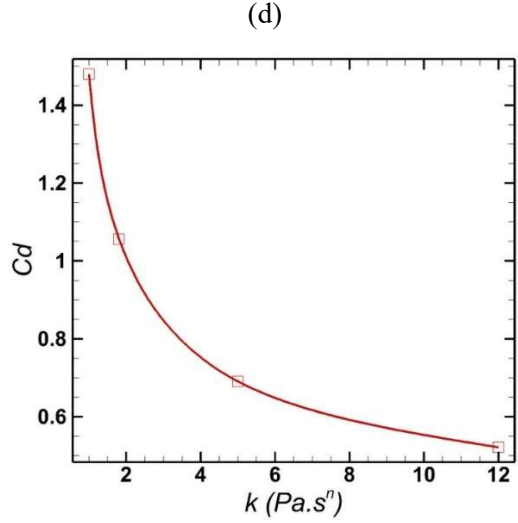

Figure 11. a) Effects of the consistency constant $k$ on the yielded (red) and unyielded (blue) zones. b) Variation of stress magnitude along $x$– axis at $y = 0$ and c) dependence of characteristic lengths on $k$. d) Drag coefficient on the cylinder versus $k$. The rest of parameters are those of fluid 2, Table 2.

In Fig. 11 the effect of the consistency index $k$ on the yielded/unyielded regions is shown. Based on their definitions, when $k$ increases, the Bingham number decreases, and the Weissenberg number increases more significantly. This can be seen quantitatively in Table 6.

| $k\ (Pa.s^n)$ | 1 | 1.81 | 5 | 12 |
|---|---|---|---|---|
| $Wi$ | 0.034 | 0.061 | 0.17 | 0.408 |
| $Bn$ | 3.375 | 1.876 | 0.682 | 0.284 |

Table 6. Effect of $k$ on $Bn$ and $Wi$

Indeed, as the consistency constant increases to $k = 12\ Pa.s^n$, the yielded regions next to the plates increase even more than they did with the increase of $n$. Similarly, $X_2$ increases faster with $k$ than with $n$, but $X_2$ increases with slightly smaller rate now (see Fig. 11(c)). The two rheological parameters have qualitatively different effects on the horizontal length of the islands, which can be barely seen at the higher $k$ value and the transition zones, which cease to exist for $k \geq 5\ Pa\ s^n$. By increasing $k$, the materials become more elastic and less plastic. Consequently, there should be a stronger overshoot in stress magnitude for higher consistency constant behind the cylinder, as shown in Fig. 11(b). However, the second maximum, which surpasses the yield stress and generates the transition zone, only appears in cases where $k = 1\ Pa.s^n$ and $k = 1.81\ Pa.s^n$. In the two cases with the highest consistency constant, viscous stresses dominate over the yield stress, the second maximum in their magnitude is not generated but $|\tau_d|$ either approaches $\tau_y$ from below or remains much smaller than $\tau_y$ (see Fig. 11(b)).

Fig. 11(d) depicts the decrease of the drag coefficient with increasing $k$, which is consistent with the material becoming more elastic and less plastic. As the material becomes more elastic, the asymmetric distribution of normal stresses around the cylinder surface becomes dominant, leading to a reduction in the drag coefficient, as in viscoelastic fluids.

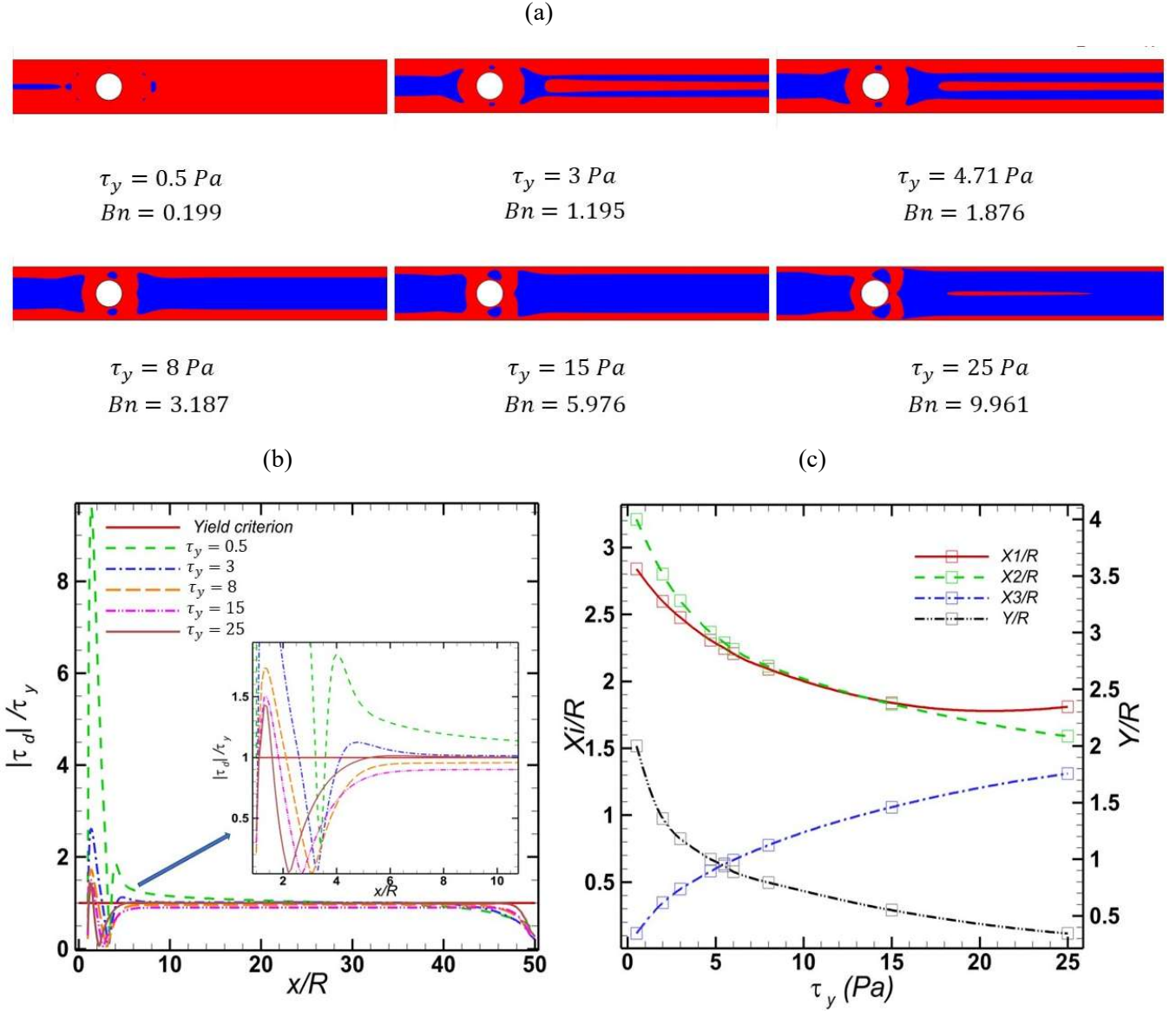

Figure 12. a) Effects of the yield stress $\tau_y$ on the yielded (red) and unyielded (blue) zones. b) Variation of stress magnitude along $x$-axis at $y = 0$ and c) dependence of characteristic lengths on $\tau_y$. The rest of parameters are those of fluid 2, Table 2.

Fig. 12 (a) illustrates the development of the yielded and unyielded regions, when $\tau_y$ increases, which proportionally increases only the $Bn$. For the smallest value of the yield stress, $\tau_y = 0.5\ Pa$, the yielded regions prevail almost throughout, with only three small islands observed at the front and three at the rear of the cylinder, a small unyielded central region upstream of the cylinder and no such region downstream of it, at least up to the time we computed the solution. When $\tau_y$ is increased, the material solidifies as before. For $\tau_y = 3\ Pa$ and $4.71\ Pa$, transition zones appear, the thickness of which decreases as the yield stress increases. For the next pair of values reported, $\tau_y = 8\ Pa$ and $15\ Pa$ they do not arise but reappear when $\tau_y = 25\ Pa$. Their initial appearance followed by disappearance is easily explained, considering that for the first pair of $Bn$ the second maximum of stress magnitude surpasses the yield stress, something that is not achieved for the next pair of $Bn$ values (see Fig 12(b)). For the highest reported $Bn$ value, the transition zone is different from before because it extends to a considerably smaller distance, resembling very much the transition zone reported in Fig. 9 of [23]. It seems that for $\tau_y = 25\ Pa$, the two islands have increased abruptly and nearly block the constrictions between the cylinder and the plates. This increases the main velocity and all stress components, so that the stress magnitude manages to overcome slightly the increased yield stress. However, this increase is very fast nullified by the high value of $\tau_y$. According to Fig. 12(b), for $\tau_y = 25\ Pa$, the stress magnitude exhibits a weak second maximum, primarily caused by axial elongation. This overshoot surpasses the yield stress for a short length, but ultimately, the high value of the yield stress overcomes the stress magnitude. This constitutes another example that by increasing the yield stress, elastic effects are increased as well.

The above observations are quantified in Fig. 12 (c), where results with several additional values of $\tau_y$ are included. As expected, when $\tau_y$ increases, the length of the islands increases, the two unyielded regions in the upstream and downstream channels get closer to the cylinder, and the yielded area next to the downstream wall gets narrower, nearly approaching zero, but not allowed to reach it because the flow rate must be maintained. All these variations are monotonic. Only the thickness of the transition zone varies non-monotonically, for reasons explained above.

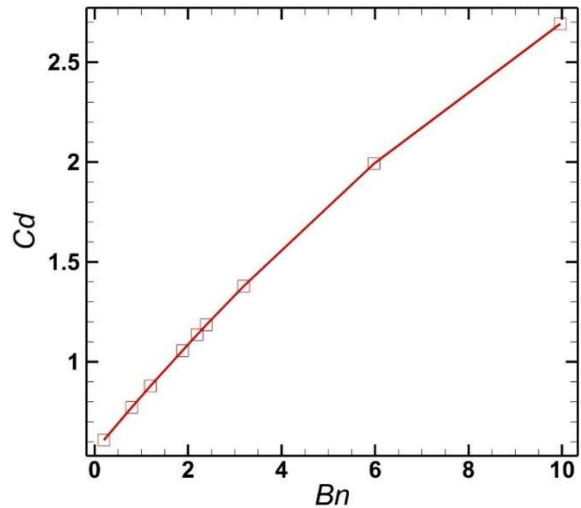

Figure 13. Drag coefficient around the surface of the cylinder versus the Bingham number.

Fig. 13 illustrates the dependence of the drag coefficient on the Bingham number. In the range of $0.1 \leq Bn \leq 3$ the drag coefficient grows almost linearly. This growth is caused by the narrowing of the yielded material between the cylinder and the plates, which leads to intensification of the primary velocity, resulting in intensification of all stress components and, naturally, an increase in the drag coefficient. For higher values of $Bn$ the increase of the drag coefficient is slower.

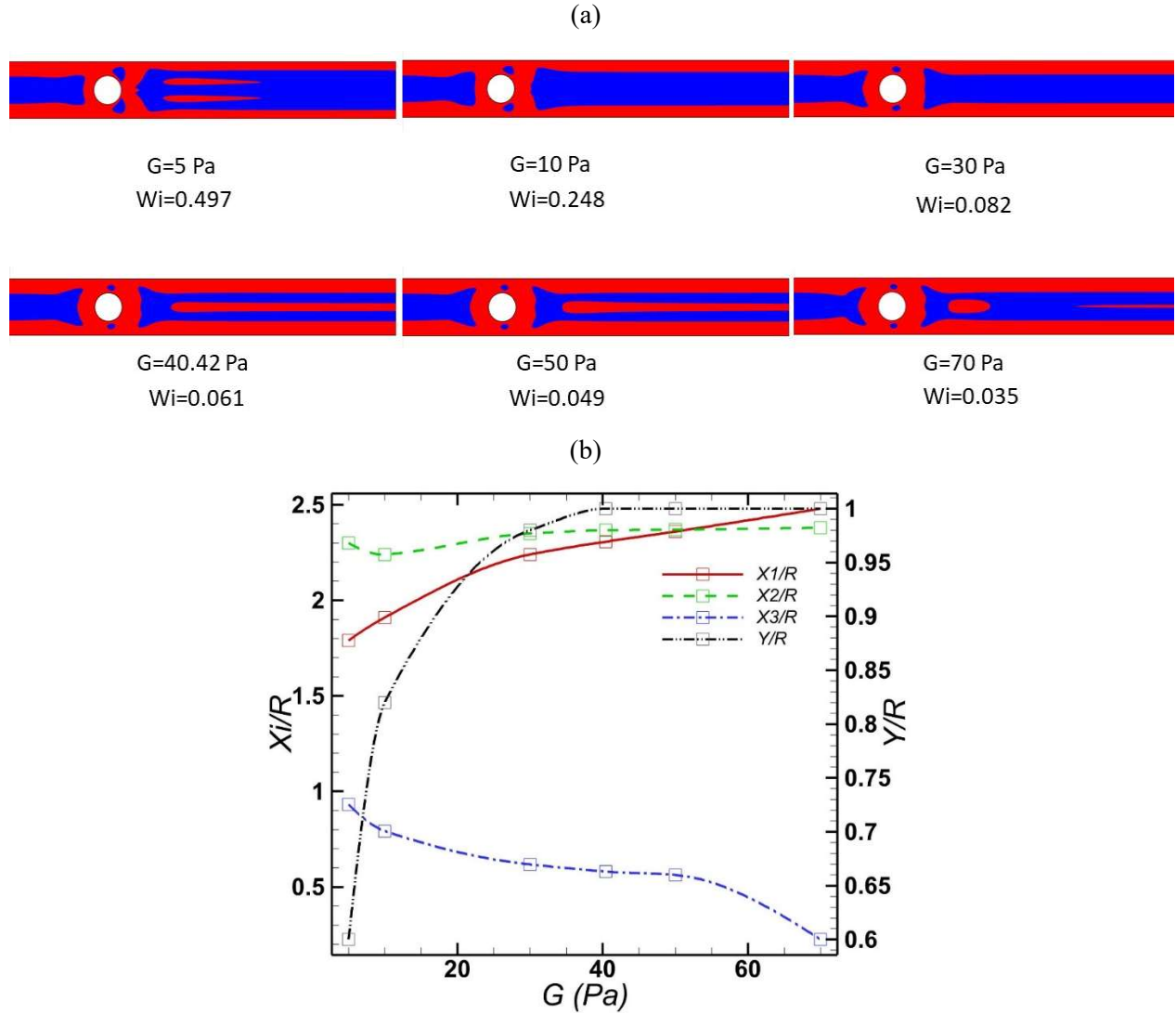

Figure 14. a) Effects of the elastic modulus $G$ on the yielded (red) and unyielded (blue) zones. b) Dependence of characteristic lengths on $G$. The rest of the parameters are those of fluid 2 in Table 2.

Fig. 14 (a) shows the effect of the elastic modulus $G$ on the yielded/unyielded regions. The elastic modulus solely affects the Weissenberg number and is inversely proportional to it. As materials become more elastic, they deform more prior to yielding. When elasticity dominates and a material is deformed it may rebound. Conversely, stiffer materials yield more readily after smaller deformation and are less able to revert to their original shape. As $G$ decreases, the EVP materials become more elastic, resulting in an expansion of unyielded areas [32, 45]. Conversely, higher values of $G$ make the material stiffer and more prone to yielding, explaining the appearance of the yielded area along the centerline in the downstream channel. Additionally, higher elasticity increases the size of the islands and relocates them in the flow direction.

Interestingly, when $G = 5\ Pa$, the emergence of the strongest negative wake behind the cylinder significantly affects yielding transitions, leading to material fluidization not predicted heretofore: two transition zones of finite length arise and are symmetrically placed with unyielded material between them.

These are caused by significant elongation in both the $x$- and $y$-directions, because both $\tau_{xx}$ and $\tau_{yy}$ contribute significantly to material yielding. When the elastic modulus increases from its lowest values to $G = 10\ Pa$ and $G = 30\ Pa$, the transition zones disappear just like they did when $\tau_y$ decreased from its highest value, and for the same reason. Similarly, further increase of $G$ to $G = 40.42\ Pa, G = 50\ Pa$ and $G = 70\ Pa$, reinstates the transition zones, following the pattern of decreasing $\tau_y$, except for the highest value with the stiffest material, where the transition zone splits into two areas. For $G = 70\ Pa$, the stress magnitude exhibits a strong second maximum, causing the very stiff material to take some time to revert to its original shape. As a result, the transition zone does not appear continuously (see Fig. 14(a)).

Fig. 14 (b) quantifies the above observations and introduces additional data points. The horizontal distance of the tip of the upstream unyielded zone ($X_1/R$) increases with an increase in $G$. On the other hand, the horizontal distance of the tip of the downstream unyielded region ($X_2/R$) experiences a small fluctuation, stabilizing after reaching a $G = 40\ Pa$. This non-monotonic behavior is caused by the penetration of the yielded region inside the unyielded one for $G = 5\ Pa$ and $G = 10\ Pa$, which ceases to exist for higher $G$ values. This transition of the tip shape from concave to convex occurs in highly elastic materials and will be encountered again in the subsequent section where transient flows are discussed. As expected, the thickness of the yielded region downstream from the cylinder increases by increasing $G$ and the horizontal length of the islands decreases. It appears that for $30\ Pa \leq G \leq 50\ Pa$, the length of the islands remains relatively constant and then decreases again until $G = 70\ Pa$.

One of the reviewers pointed to us that in other studies, namely [46, 47] it has been reported that the yielded area next to a wall increases when fluid elasticity increases. However, in [46] the augmented Lagrangian method was used to directly solve for the steady Poiseuille and Couette flows, whereas it is known that only transient simulations must be used with EVP materials even when a steady flow is sought [10, 48]. For the same reason parameter continuation cannot be used in EVP simulations to obtain a new steady state solution from a previous one, because plastic stresses from the first solution will affect the next solution, leading to irrelevant results. Moreover, Izbassarov et al. [47] partly examined laminar channel flow, but used a ratio of the solvent to the total viscosity $\beta = 0.9$, which means that the material is primarily Newtonian and, hence, elastic and plastic effects are secondary. Clearly, a more elastic solid may undergo greater deformation before yielding. Upon further increase of the applied force it will yield creating a smaller yielded area, through which more intense fluid flow will take place because of the reduced cross section. In contrast, stiffer materials yield easier after smaller deformation and provide a wider cross section for the flow, which, hence, is less intense. This is a general outcome from several studies by several research groups, [45, 49, 50] particularly when the flow is pressure or stress driven. Things may become more complicated when the flow is buoyancy driven.

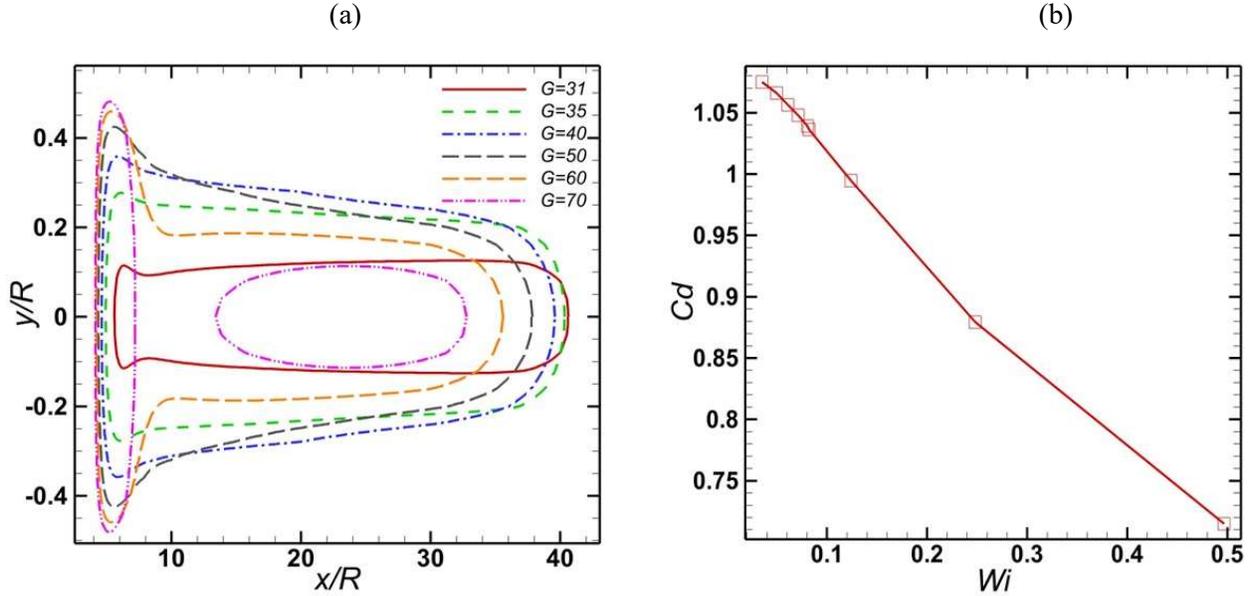

Figure 15. a) Illustration of the location of yield surface for different elastic modulus and b) drag coefficient around the surface of the cylinder versus the Weissenberg number.

Fig. 15 (a) depicts the dependence of the location and shape of the transition zones on the elastic modulus. For $G = 30\ Pa$, no transition zone exists but increasing it by just $1\ Pa$, a transition zone starts to appear. Therefore, there exists a critical value of $Wi$, above which transition zones appear. For fluid 2 and $B_R = 0.5$, this critical value is 0.082. The length of these yielded areas tends to decrease as $G$ increases. In particular, their downstream boundary tends to move away from the outflow boundary, while their upstream boundary translates less but still away from the cylinder. This decrease in length as elasticity decreases is a natural consequence of the development of lower stress magnitude, which surpasses the yield stress to a smaller degree, allowing them to remain in the transition zone for shorter distance. Conversely, the thickness of the transition zone, especially near the cylinder, significantly decreases as the elastic modulus decreases. Further away from the cylinder they tend to also contract, but not monotonically.

Fig. 15 (b) demonstrates that increasing material elasticity decreases the drag coefficient on the cylinder. A higher $Wi$ indicates materials with greater elasticity and higher resistance to yielding. This decrease in drag for small $Wi$ has been reported repeatedly in the literature, albeit for viscoelastic liquids; see also Fig. 24 (b). In viscoelastic fluids, it has been attributed to an asymmetry in the variation of normal stress components around the surface of the cylinder [51]. This asymmetry is also observed in EVP materials (see Fig. 2 (b)) and becomes more pronounced along the surface of the cylinder, as the Weissenberg number increases. As $Wi$ increases, the position of the maximum $\tau_{xx}$ shifts to the right surface of the cylinder (data not presented here for conciseness), which also increases the pressure in this region which results in a decrease in drag. Moreover, the shear thinning effect also contributes to the reduction of drag, as reported in [51].

## 3.5 Unsteady flow when G decreases below a critical value

When we extended our study to even lower values of the elastic modulus or larger values of the yield stress, we observed that the flow remained transient or led to complicated patterns. The time-dependent behavior in EVP materials can emerge at lower elasticity levels (lower $Wi$) compared to viscoelastic fluids, as noted in previous studies [32, 45]. Hence, we consider it important to dedicate a separate section to this phenomenon. Since the resulting velocity and stress fields are quite more complicated, the refined mesh, D5, given in Table 3, was used in all these simulations. The related mesh convergence test is given in the Appendix D. All simulations were carried out until $t = 200\,s$. First, we discuss the effect of decreasing $G$. As the elastic modulus decreases, the Weissenberg number increases, leading to elastic instability both upstream and downstream the cylinder. This transient response of the EVP material intensifies as the elastic modulus approaches unity.

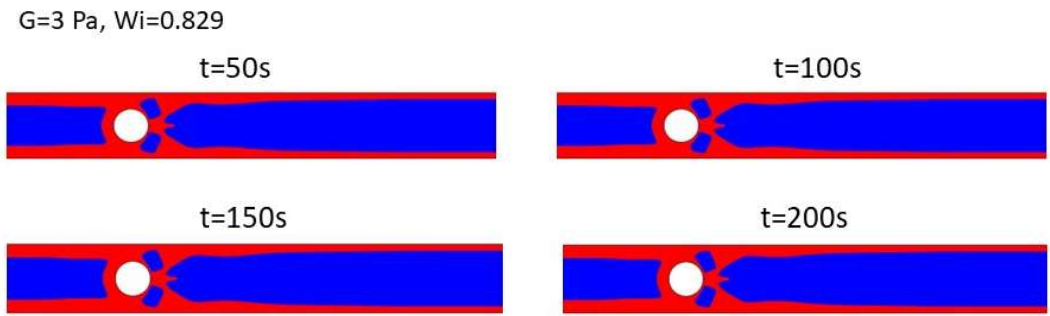

Figure 16. Evolution of the yielded (red) and unyielded (blue) zones for $G = 3\,Pa$ ($Wi = 0.829$), while the rest of the parameters are those of the fluid 2 in Table 2. The yield strain is $\varepsilon_y = 1.57$.

Fig. 16 illustrates four snapshots of the yielded and unyielded regions around the cylinder for $G = 3\,Pa$. The video in the supplementary material is even more revealing. Small oscillations are initially observed in the early snapshots, but they completely damp out later and the yielded and unyielded areas reach a steady state. Yielded material penetrates a little inside the centrally located unyielded region downstream of the cylinder, something observed only for low $G$ values and forms an indentation. The downstream unyielded area slowly expands up to an axial position of about $\sim 10R$, but stabilizes thereafter. The islands' shape changes from a circular to a more rectangular one. Additionally, these islands tend to connect to the main unyielded region downstream of the cylinder, which indicates once more that elasticity relocates the unyielded regions in the flow direction.

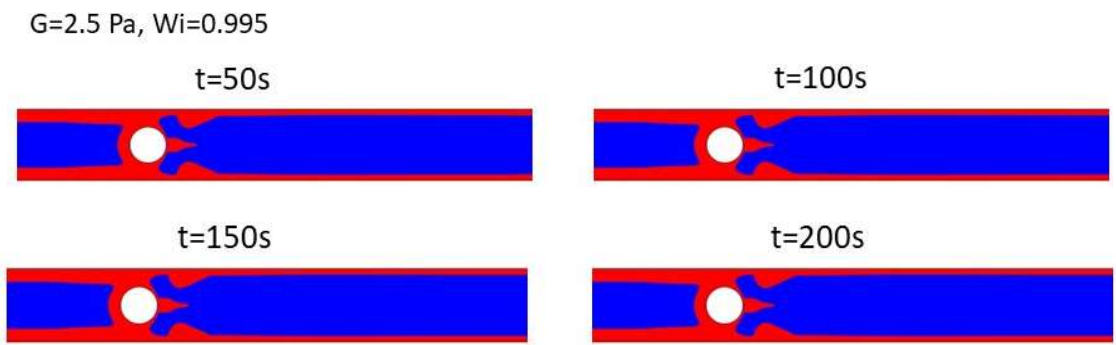

Figure 17. Snapshots of the yielded (red) and unyielded (blue) zones for $G = 2.5\,Pa$ ($Wi = 0.995$), the rest of the parameters are those of the fluid 2 in Table 2. The yield strain is $\varepsilon_y = 1.884$.

When $G$ decreases to $G = 2.5\ Pa$, as shown in Fig. 17, the same patterns emerge as in $G = 3\ Pa$, which eventually stabilize, achieving a steady state. This can be better understood by watching the corresponding video in the supplementary material. First, immediately after the cylinder, a burst of a wave package with 4-5 maxima is generated and translates downstream. Such a package cannot be observed in the snapshots, as it occurs either before or after the corresponding frames shown in Fig. 17. Eventually, the irregular variations of the yield surface stabilize throughout the downstream region of the cylinder. The main unyielded region upstream from the cylinder displays slight vibrations that damp out over time completely. Furthermore, the islands at the top and bottom of the cylinder are connected to the main unyielded region downstream. Behind the cylinder, a polar cap can be seen, followed by the acute penetration of yielded material into the main unyielded region. The symmetry plane is maintained throughout all the snapshots shown.

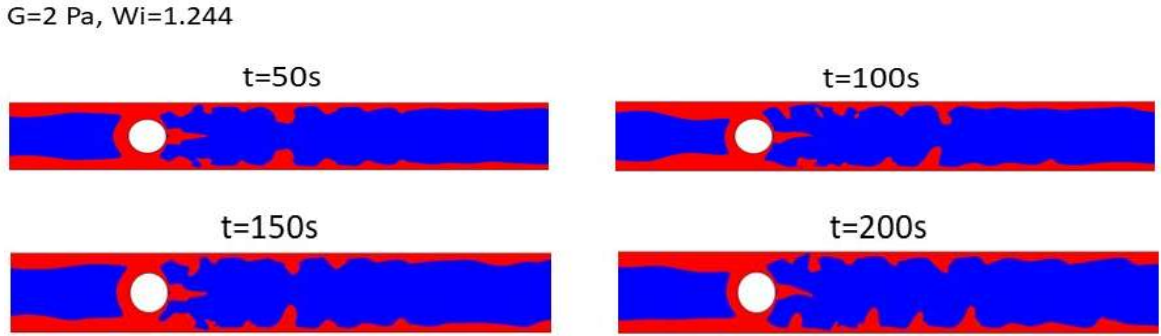

Figure 18. Snapshots of the yielded (red) and unyielded (blue) zones for $G = 2\ Pa$ ($Wi = 1.244$), the rest of the parameters are those of the fluid 2 in Table 2. The yield strain is $\varepsilon_y = 2.355$.

The pattern described above is further intensified when the elastic modulus decreases further to $G = 2\ Pa$, as better seen in the SM. Immediately, a wave package emerges very near the cylinder and translates downstream leaving behind nearly undisturbed the yield surface. This is followed by another similar event, but afterwards the generated wave package remains close to the cylinder and damps out towards the end of the domains depicted in Fig. 18. This can be seen in all snapshots, i.e., this pattern persists over time. These waves exhibit completely irregular three-dimensional patterns, with their shapes changing over both time and space. Additionally, the entire unyielded zone upstream of the cylinder undergoes oscillations. Its upper and lower yield surface may oscillate in phase (sinuous mode) or out of phase (varicose mode) and its end near the cylinder may undergo irregular oscillations. The corresponding video in SM shows that these oscillations are intermittent. The two islands have merged again with the downstream unyielded area and created an acute indentation of yielded material in its front. It is worth noting that in this case with the lowest value of $G$ that we examined, the unyielded regions are closest to the cylinder and seem to encompass the entire surrounding area.

$t = 50\ s$

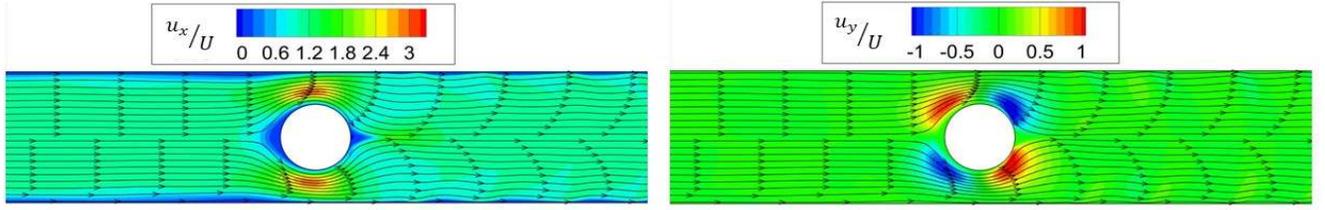

$t = 100\ s$

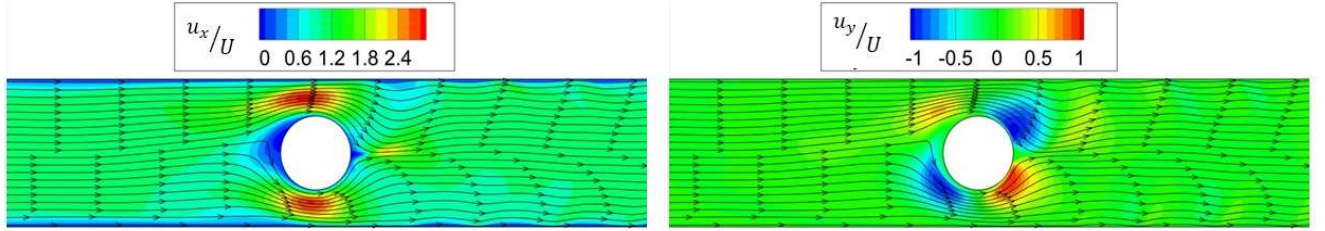

$t = 150\ s$

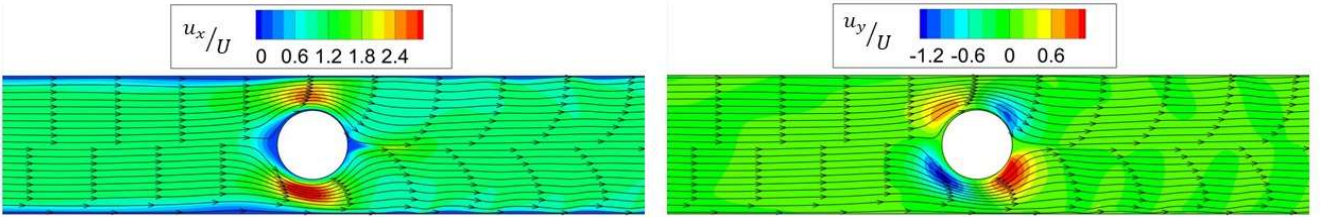

$t = 200\ s$

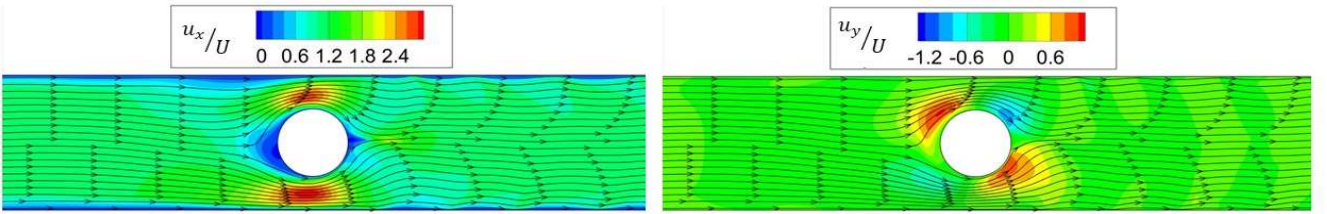

(a)            (b)

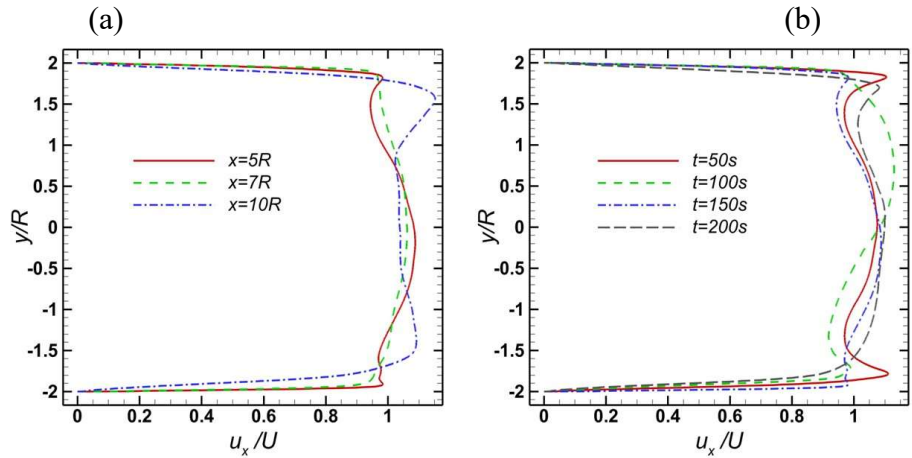

(c) (d)

Figure 19. Evolution of $(a)\frac{u_x}{U}$, and $(b)\frac{u_y}{U}$, for $G = 2\ Pa$, while the rest of the parameters are those of the fluid 2, in Table 2. Variation of dimensionless velocity $(\frac{u_x}{U})$ at c) $t = 150s$ for three $x$-locations d) $x = 5R$ for four time-instants.

In Fig. 19 the velocity components and path lines are illustrated for $G = 2\ Pa$ and at the same time instants discussed above. In the supplementary material the stress components, and pressure field are given for the same time instants. Asymmetric flow between the top and bottom of the cylinder can be observed in all snapshots. At $t = 50\ s$, the path lines are relatively evenly distributed between the upper and lower parts of the cylinder. The maximum $u_x$ is located at the top and bottom of the cylinder. The secondary velocity reaches its maximum in the northwest and southeast positions and its minimum in the northeast and southwest positions around the cylinder, as expected. Fluctuations in the horizontal velocity and the path lines are evident behind the cylinder and intensify in the yielded region of the material. Moving to $t = 100\ s$, the path lines bend slightly downward just before the cylinder. However, a clear asymmetry in the lateral velocity is observed between the top and bottom surfaces. At $t = 150\ s$, the path lines mainly move to the lower part of the cylinder, indicating that the fluid is predominantly flowing from the bottom opening. The vertical velocity remains similar to the previous cases in all snapshots until $t = 150\ s$. Finally, at $t = 200\ s$, the path lines exhibit more amplified waves both upstream and downstream the cylinder. Here, most of the fluid flows through the lower opening. Furthermore, the vertical velocity differs from the previous snapshots, with the maximum $u_y$ in the southeast and less dominant in the northwest. It is worth noting that from the initial snapshot to the final one, no plane of symmetry is observed either in $u_x$ or $u_y$, in agreement with the discussion related to Fig. 19. Clearly these variations in the velocity are smaller than the variations seen in the yield surfaces (Fig. 19) which in turn were caused by variations in the stress field.

A nearly stagnant region appears in the upstream channel at $t = 150\ s\ and\ t = 200\ s$, similar to the findings of Hopkins et al. [52] for viscoelastic fluids. As the Weissenberg number ($Wi$) increases, the flow experiences elastic instability, starting with the bending of streamlines near the cylinder. This flow causes the emergence of small stagnant regions on the channel walls. The appearance of a stagnant region is affected by the fluid's elasticity, shear-thinning, and geometrical factors such as the blockage ratio [52]. This phenomenon appears at much higher $Wi$ in viscoelastic fluids [52] compared to our prediction for EVP fluids due to the presence of plasticity.

It should be noted that the primary velocity does vary even in the unyielded region, as seen in Figs. 19(c) and 19(d) both in time and $x$-location, something that is not clearly visible in Figs. 19(a) and 19(b). This indicates that the material within the unyielded regions is also deforming. Consequently, in addition to the strong variation of the normal elastic stress components, the velocity field does vary here.

As stated in [32], a purely elastic and inertialess flow instability, arises from the interaction of curved path lines and high normal elastic stresses even in EVP fluids. The original instability mechanism proposed by McKinley and co-workers [53, 54] examined steady viscoelastic flows, whereas here the flow is inherently transient and remains so. Nevertheless, the criterion seems to apply qualitatively here as well. On the contrary, the disturbances intensify over time. It is important to note also that purely viscoelastic instabilities have been reported to emerge in the spanwise direction ($z$-axis) once a critical Weissenberg number, $Wi = 1.3$, is surpassed, a value higher than the maximum Weissenberg number reported here [55].

Producing EVP fluids with so high elasticity may seem difficult. Nevertheless, very recently Abdelgawad et al. [56], but also by Sen et al. [57] combined Pluronic or Carbopol, respectively, with PEO and increased the extensional properties, not the shear properties of the respective EVP fluid. The base material used in our study was plain Carbopol and was characterized in shear experiments only (flow curve and SAOS). Certainly, these two papers open new possibilities. The two new materials may require a modification of the SHB model to account independently for their very different extensional properties. Increasing the extensional viscosity alone may facilitate the experimental observation of the time dependent flows we predict when elasticity increases.

## 3.6  Unsteady flow when $\tau_y$ increases above a critical value

Next, we examine the effect of increasing the yield stress. We begin with $\tau_y = 33\ Pa$ in Fig. 20. We observe that even at $t = 200\ s$ the flow retains its plane of symmetry, except near the downstream pole of the cylinder, and remains slightly transient. This is more clearly seen in the corresponding video in SM. There one can observe that all yield surfaces including those upstream from the cylinder and the transition zone undergo small amplitude oscillations. The amplitude of these oscillations is larger closer to the cylinder, and gradually decreases downstream to the point that at $x \approx 15R$ it is not visible.

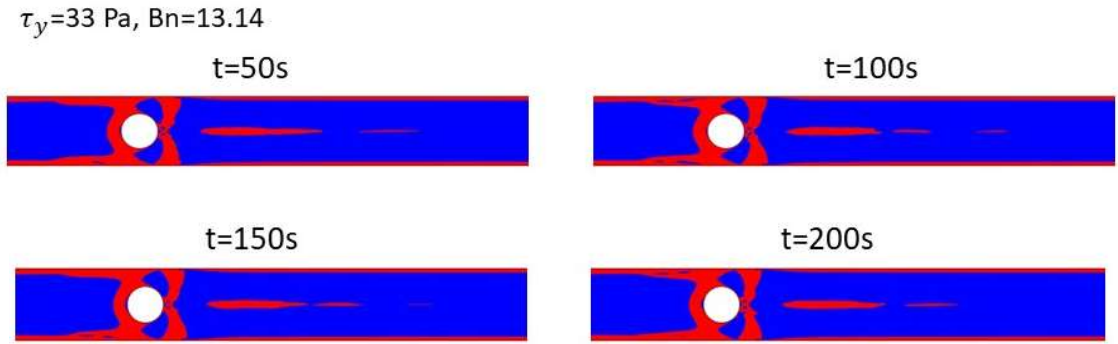

Figure 20. Snapshots of the yielded (red) and unyielded (blue) zones for $\tau_y = 33\ Pa\ (Bn = 13.14)$, the rest of the parameters are those of the fluid 2 in Table 2. The yield strain is $\varepsilon_y = 0.816$.

While these oscillations do not have a high amplitude, they influence the position of the tip of the two unyielded regions. Upon closer examination, we observe that even the front polar cap undergoes expansion and contraction over time, which is due to a rapid variation of the stress field in its vicinity. Between the cylinder and the islands, which also undergo fluctuations, a thin yielded film emerges due to intense local shearing. As we will see next, when the yield stress increases, the thickness of this thin film decreases. To establish the yielded material in this region, the local stress field is large, leading to the detachment of unyielded material from the islands. This material forms typically two small unyielded regions just behind the rear polar cap, with fluctuating sizes.

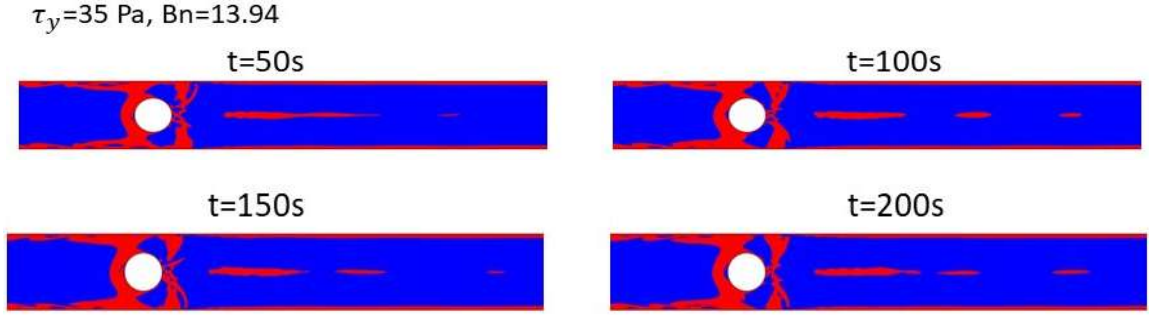

Figure 21. Snapshots of the yielded (red) and unyielded (blue) zones for $\tau_y = 35\ Pa$ $(Bn = 13.94)$, the rest of the parameters are those of the fluid 2 in Table 2. The yield strain is $\varepsilon_y = 0.865$.

When the yield stress increases to $\tau_y = 35\ Pa$, all fluctuations become more intense and more complex, although they remain near the cylinder (see Fig. 21). This intensification includes the unyielded area upstream of the cylinder, and the transition zone, which not only has a fluctuating length as before, but also a wavy yield surface. The downstream yielded area is the only part of the domain that quickly reaches a steady thickness. The two islands gradually shift in the flow direction, and undergo stronger shape and volume oscillations, leading to the detachment of two or more small unyielded regions near the rear pole, which fluctuate as well in both size and shape. Finger-like yielded zones are formed in the main unyielded area, but they are not symmetric, instead they fluctuate asymmetrically. Such fingers do not arise in the upstream unyielded area. These instabilities are more clearly visible in the corresponding video in SM.

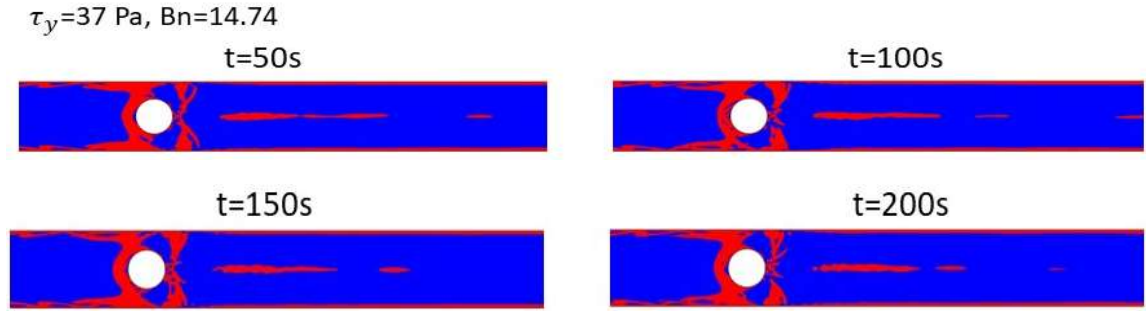

Figure 22. Snapshots of the yielded (red) and unyielded (blue) zones for $\tau_y = 37\ Pa$ $(Bn = 14.74)$, the rest of the parameters are those of the fluid 2 in Table 2. The yield strain is $\varepsilon_y = 0.915$.

The largest yield stress value we examined is $\tau_y = 37\ Pa$, and the corresponding snapshots can be seen in Fig. 22. Fluctuations of all yield surfaces described above intensify further and some of them become erratic as depicted more clearly in the corresponding video in SM. The two islands increase in size, and approach the downstream unyielded area, while unyielded material still gets detached from them. The main new feature in this case is the stronger fluctuation of the thin yielded film upstream of the cylinder and next to the wall, where stronger elastic stresses develop, leading to the formation of finger-like asymmetric unyielded islands that translate downstream and eventually disappear. Starting from $t = 50\ s$, several small unyielded regions emerge near the islands, elongating towards the downstream channel and eventually connecting to the main downstream unyielded region. Simultaneously, the polar caps in both upstream and downstream areas grow in size. The transition zone is consistently present in all snapshots, oscillating over time. The plane of the symmetry is broken as in both previous cases.

An interesting difference between high Bingham and high Weissenberg number flows is that in the former case the islands remain detached from the main unyielded region downstream from the cylinder, whereas in the latter cases they merge with it, resulting in the formation of a larger unyielded zone. It appears that elasticity plays an important role in the elongation of the unyielded regions in the direction of the flow, thereby linking the islands to the main unyielded region in the downstream channel.

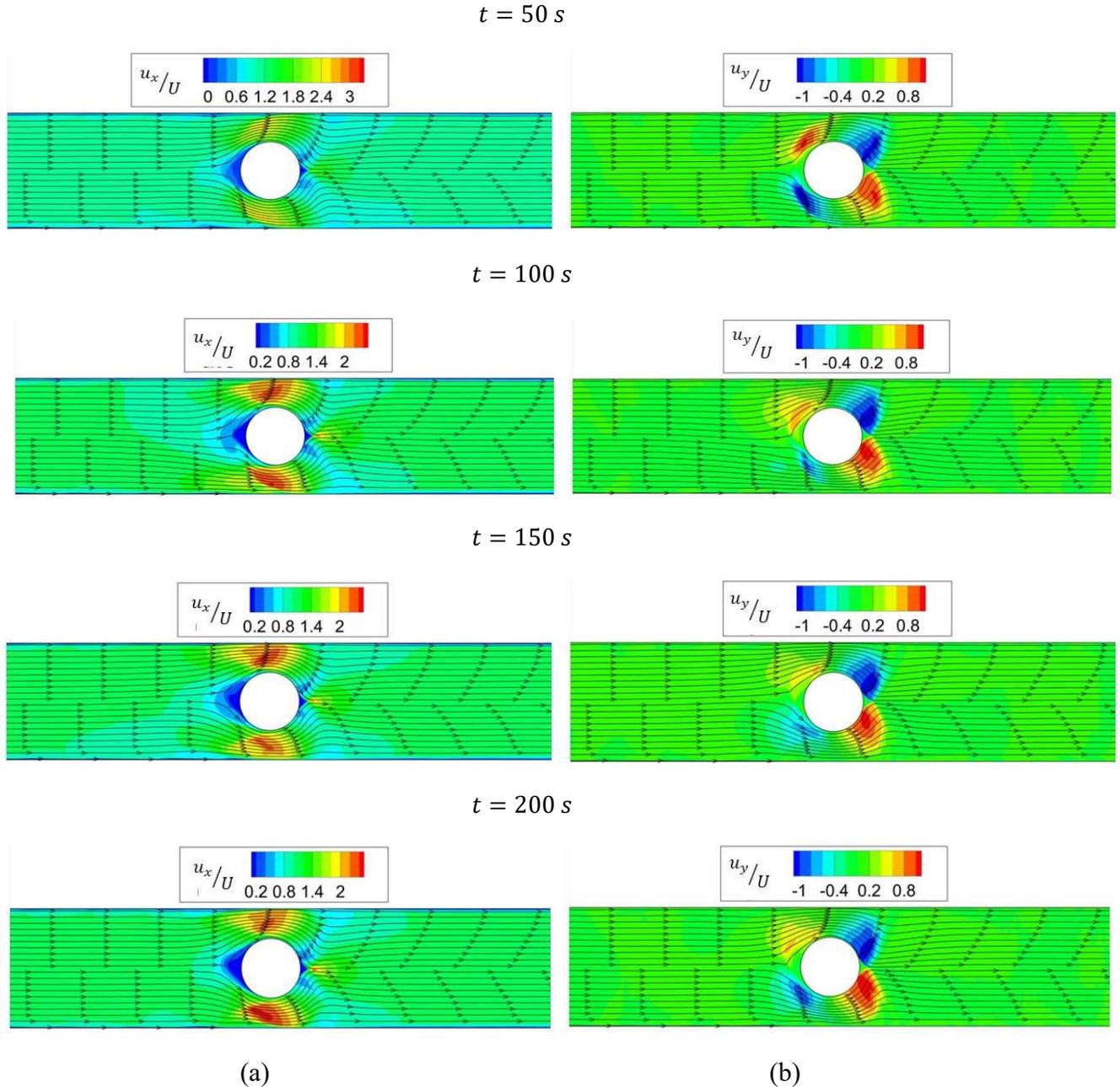

Figure 23. Evolution of $(a)\ \frac{u_x}{U}$, and $(b)\ \frac{u_y}{U}$, for $\tau_y = 37\ Pa$, while the rest of the parameters are those of the fluid 2, in Table 2.

Fig. 23 illustrates the horizontal and vertical velocity fields for $\tau_y = 37\ Pa$. At $t = 50\ s$, the maximum $u_x$ appears at the top of the cylinder, with the minimum $u_x$ occurring near the front and back stagnation points. Close to the cylinder, four areas of minimum and maximum values of $u_y$ can be seen. The location of these extreme values of $u_y$ are asymmetric in all snapshots. The path lines exhibit uniform distribution and are less curved compared to the high $Wi$ cases (see Fig. 19). Moving to $t = 100\ s$, there are regions with high $u_x$ above and below the cylinder. At $t = 150\ s$, the path lines indicate that the fluid moves predominantly above the cylinder, which concurs with the maximum of $u_x$ appearing there. Finally, at $t = 200\ s$, the path lines bend slightly more towards the bottom of the cylinder, and the maximum $u_x$ is more expanded in this region. The maximum horizontal velocity is slightly higher in the high $Wi$ flows compared to high $Bn$ flows (compare with Fig. 19). The corresponding stress and pressure fields are illustrated in the supplementary material.

As discussed also in [32], the mechanism leading to oscillatory or erratic stress fields and yield surfaces when $Bn$ increases is similar to the mechanism leading to similar phenomena when $Wi$ increases. Only the sequence of events changes. Here, when $Bn$ increases the yielded film between the islands and the cylinder becomes thinner, promoting the development of stronger elastic forces, which lead to the detachment of small unyielded materials. This effect can be seen in other areas of the domain, finally leading to stronger elastic stress components. Hence, this requirement of the purely elastic instability increases indirectly, and its coupling with the curved path lines generates it [53, 54]. As stated also in [32], the instability is more intense when the elastic modulus decreases, as opposed to when the yield stress increases.

| $\tau_y\ (Pa)$ | 8 | 15 | 25 | 33 | 35 | 37 |
|---|---|---|---|---|---|---|
| $X_1/R$ | 2.092 | 1.84 | 1.81 | 1.722-1.772 | 1.712-1.773 | 1.7-1.773 |
| $X_2/R$ | 2.112 | 1.83 | 1.59 | 1.485-1.493 | 1.466-1.499 | 1.431-1.504 |
| $Y/R$ | 0.794 | 0.55 | 0.344 | 0.213 | 0.195-0.199 | 0.174-0.193 |
| $\varepsilon_y = \tau_y/(G = 40.42\ Pa)$ | 0.197 | 0.371 | 0.618 | 0.816 | 0.865 | 0.915 |
| $G\ (Pa)$ | 2 | 2.5 | 3 | 5 | 10 | 30 |
| $X_1/R$ | 1.683-1.7 | 1.71 | 1.69 | 1.79 | 1.91 | 2.24 |
| $X_2/R$ | 2.814-3.035 | 2.70 | 2.561 | 2.3 | 2.24 | 2.35 |
| $Y/R$ | 0.254-0.872 | 0.308 | 0.417 | 0.6 | 0.82 | 0.98 |
| $\varepsilon_y = (\tau_y = 4.71\ Pa)/G$ | 2.355 | 1.884 | 1.57 | 0.942 | 0.471 | 0.157 |

Table 7- Effect of yield stress and elastic modulus on the distance of the tip of the unyielded region in the upstream ($X_1/R$) and downstream ($X_2/R$) channels and the thickness of yielded film in the downstream channel ($Y/R$) and on the yield strain $\varepsilon_y = \frac{\tau_y}{G}$.

The main unyielded zones in both the upstream and downstream channels approach the cylinder when $\tau_y$ increases or $G$ decreases. Similarly, the thickness of the yielded film downstream of the cylinder reduces with this variation of the same parameters. Often these characteristic lengths fluctuate with time, and then an approximate range is determined for them (Table 7). The effects of the yield stress and the material elasticity on this and other flows of EVP materials can be examined more concisely by introducing the yield strain parameter, $\varepsilon_y = \tau_y/G$, which depends only on material properties and increases when either $\tau_y$ increases or $G$ decreases [32]. This dimensionless variable was first introduced in Cheddadi et al. [24] and then often used to describe phenomena in EVP fluids [23, 32]. An increase in yield strain makes the unyielded material more elastic, as manifested by the increase in the normal elastic stress components. These significantly contribute to the von Mises criterion, causing material yielding at lower shear stress values than expected. When the solvent viscosity is negligible, (as in this study with $\eta_s = 0.01\ Pas$, but is included only to retain the ellipticity of the momentum balance), the product of the Weissenberg number with the Bingham number results in the yield strain, $\varepsilon_y = \frac{\tau_y}{G} \approx Wi * Bn$. Moreover, the effect on the inception of instability by decreasing $G$ is stronger than the effect by increasing $\tau_y$, by the same increment, just because of the definition of the yield strain, $\varepsilon_y = \frac{\tau_y}{G}$, [32]. By comparing the results for high Weissenberg numbers ($Wi$) with those for high Bingham numbers ($Bn$), we concluded that yield strain alone is insufficient to determine the transition to instability.

## 4   Summary, Discussion and Conclusions

The flow around a confined cylinder was investigated using the Saramito-Herschel-Bulkley model. The governing equations were solved numerically using OpenFOAM coupled with the RheoTool toolbox, which employ the Finite volume method to discretize the equations. The accurate implementation and convergence of the numerical method was verified by examining, with successively refined meshes, the flow of viscoelastic fluids, and comparing our predictions for velocity profiles and the dependence of the drag coefficient on $Wi$ with previous studies. Moreover, our predictions favorably compare with reports by [24, 37], for elastoviscoplastic fluids.

The flow of EVP materials was investigated first examining three well-characterized Carbopol gels with concentrations of 0.09% and 0.1% (fluids 1 and 2). Surprisingly, despite their very similar concentrations, these gels displayed distinctly different yield surfaces. Among the six non-dimensional groups which affect the flow (Reynolds, Weissenberg, Bingham, yield strain, ratio of solvent to solution viscosity and blockage ratio), the Reynolds number remained much smaller than one and was set to zero and the viscosity ratio was kept constant and small.

Small unyielded areas, the so-called polar caps on the cylinder poles and the islands between the cylinder and the plates as well as around the planes of symmetry upstream and downstream of the cylinder are predicted, extending the corresponding predictions for viscoplastic fluids. Most unexpectedly, yielded material is predicted around the center plane and downstream from the cylinder under certain conditions. Local approximate analysis shows that in this region the stress magnitude decays exponentially toward the yield stress but does not fall below it. Hence the appearance of this elongated yielded area trapped inside unyielded zones is what has been called "transition zone". Its appearance depends on the interplay of fluid elasticity, yield stress, and blockage ratio.

An important factor affecting the results is the location of the downstream boundary, where the usual boundary conditions have been applied. It was found that this location does affect the flow and particularly the transition zone, generating a non-acceptable solution for a length of around $10R$ ahead of the outflow boundary. Because of this, in all solutions reported we have ignored the flow in this region. By increasing the blockage ratio, the shear between the cylinder and the plates, and the negative wake intensify, while the transition of the normal stresses is more abrupt. All these lead to an increase in the drag coefficient with $B_R$. The transition zone appears for $B_R > 0.3$.

We expanded our study by examining the effect of each one of the four rheological parameters ($\tau_y, G, n, k$) of the SHB model one at a time, while keeping the rest of the values to those of one of fluid 2. The numerical results are presented mainly in terms of the yield surfaces. Their shapes are characterized by the four lengths defined in Fig. 10. Both $n$ and $k$ affect four of the dimensionless groups. Their more significant effect on the flow is that the thickness of the yielded film next to the wall downstream of the cylinder increases, when $n$ or $k$ increase. By definition, $\tau_y$ and $G$ affect only $Bn$ and $Wi$, respectively. When $G$ increases, the elasticity level decreases, resulting in stiffer materials, larger yielded areas, and larger drag coefficient. Moreover, the transition zone appears only for $G > 30$ or $Wi < 0.082$. Increasing $\tau_y$ increases the unyielded areas everywhere, including the size of the islands and polar caps around the cylinder, and, naturally, the drag coefficient.

All the results discussed so far were obtained when a steady state was reached and when the Bingham and Weissenberg numbers remained below a certain critical value. When their respective critical value was exceeded, the flow remained transient or became periodic without regular wavelength or amplitude patterns, which were predicted also in [32]. The mechanism leading to this elastic instability, when $Wi$ increased, should be related to the coupling of the increased elastic stresses with the instantaneous radius of curvature of the path lines, which implies an extension of the criterion first reported in [53, 54] for steady flows. A similar instability when $Bn$ is increased is quite unexpected, because under these circumstances the flow should tend to become slower, smoother and hence steadier. A closer examination revealed that increasing $Bn$ decreases the thickness of the yielded film next to the cylinder surface, which in turn increases the elastic stresses and disrupts the nearby islands by inducing small finger-like unyielded regions around them.

A more detailed examination of the transition zone is essential for future research, especially for other EVP materials. These calculations should be compared with experimental measurements to understand the transition zones better. The persistent transient and erratic flows when $Bn$ and $Wi$ exceed certain critical values are very interesting and need to be investigated experimentally as well. Moreover, this study can be extended to an array of cylinders, an arrangement resembling flow in porous media, positioned at specific distances from each other to examine how these elastic/plastic instabilities affect the flow field between the cylinders [5].

**Appendix A:** Validation of numerical implementation of OpenFOAM with viscoelastic fluids

In this section, $\beta$ is assumed to be 0.59 to compare our results with those of previous studies. The Weissenberg number varies from 0 (Newtonian fluid) to 10. The inertia terms are neglected in all simulations. High spatial resolution is generated especially near the stagnation points to more accurately capture the flow structure there (e.g. the negative wake). The mesh is refined also around the cylinder surface, while the mesh resolution decreases as the inlet and outlet boundaries are approached. The number of cells that are used in three different meshes along with other characteristics are given in Table 8. Mesh C3 is used for all simulations with viscoelastic fluids.

| Mesh | Number of cells | Number of points | Number of cells on the surface of the cylinder | Number of y-cells at any cross section away from the cylinder |
|---|---|---|---|---|
| C1 | 14400 | 29520 | 280 | 40 |
| C2 | 24000 | 50710 | 378 | 120 |
| C3 | 30000 | 61000 | 400 | 100 |

Table 8- Characteristics of the meshes used in the viscoelastic flow around the confined cylinder.

First, we examine the velocity profile to ensure that the fully developed condition has been attained well before the cylinder. Furthermore, the drag coefficient on the cylinder is also used to verify the accuracy of our numerical scheme. Our results of the drag coefficient are compared with those in several numerical studies [58-60], and three different meshes are used to examine the effect of mesh resolution on the drag coefficient.

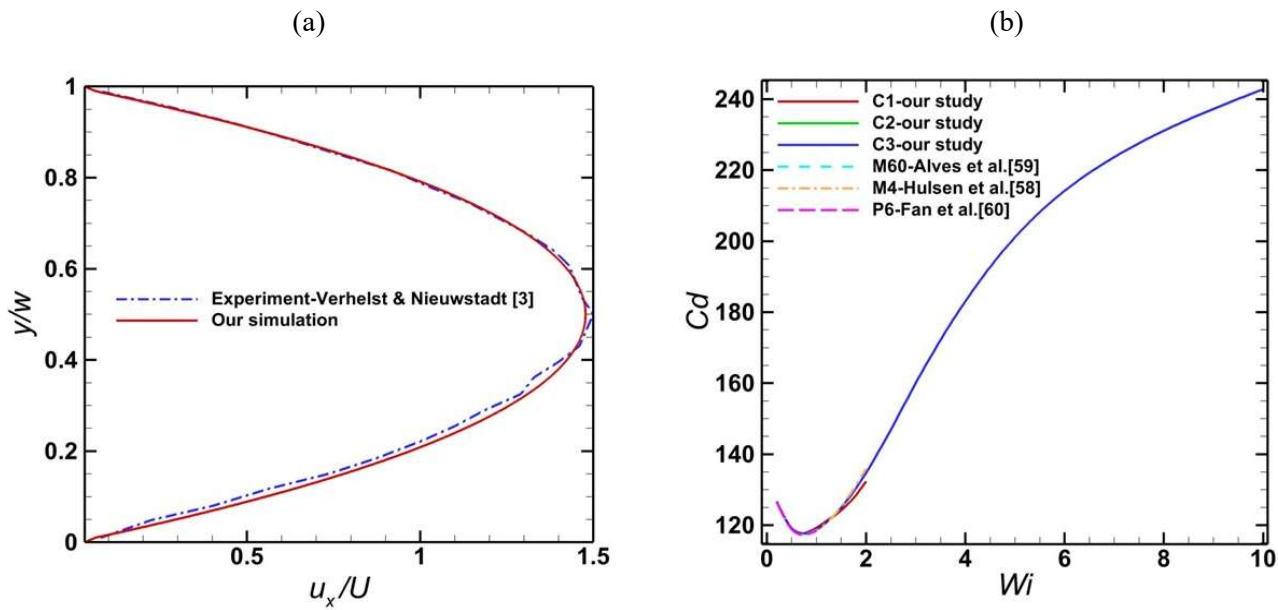

Figure 24. a) Comparison of the numerical velocity profile with the experimental measurements for viscoelastic fluid at $\frac{x}{R} = -20$ for flow around the cylinder ($Wi = 1.4, Re = 0.23$). b) Comparison of the drag coefficient of an Oldroyd-B model versus $Wi$ predicted with 3 different meshes and results from the literature.

Achieving fully developed conditions is imperative in order to eliminate any effect from the inlet conditions and, hence, isolate the influence of the cylinder. Fig. 24a depicts the velocity profile at a plane located 20 radii upstream from the cylinder. The profile is fully developed, parabolic and in agreement with in the experimentally observed by Verhelst & Nieuwstadt [3].

In Fig. 24b the drag coefficient versus the Weissenberg number for different mesh resolutions is illustrated. The simulation diverged because the mesh C1 and C2 and time step used were similar to those in [59], where it was not possible to achieve a higher $Wi$ number. Also, the results of three previous studies using Finite elements [16, 58, 60], and Alves et al. [59] using FVM are compared with ours. Mesh M60 has 17400 cells, while Mesh M4 has 4320 elements. The number of elements for Mesh P6 is not mentioned in their study, but its corresponding total number of degrees of freedom is specified as 34006 [60]. In our study, we obtained converged solutions with the highest value, $Wi = 10$, using mesh C3.

**Appendix B**: Comparison of our predictions with previous experimental and numerical reports for EVP fluids.

We compare here our predictions using OpenFOAM with experimental and numerical results in Cheddadi et al. [24, 37]. These authors examined the Saramito-Bingham model, $n = 1$, in the same flow arrangement, and reported only the dimensionless velocity at the mid-plane, both upstream and downstream from the cylinder. Moreover, they used different EVP fluid (foam) and reported its rheometric parameters, as given in the caption of Fig. 25. We performed simulations with these parameters as well. Fig. 25 shows that all three approaches capture the usual negative wake behind the cylinder and good agreement is observed, in genral. Some deviation appears at the location and magnitude of the overshoot, mainly with the predictions in [37], but less with the experiments in [24].

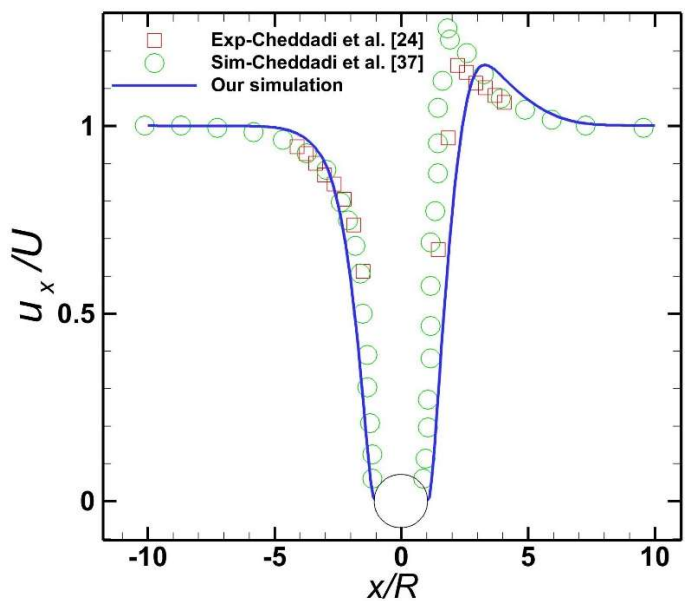

Figure 25. Comparison of the numerical velocity profile with the experimental measurements for EVP fluid (wet foam) extracted from ($B_R = 0.3, n = 1, G = 13\ Pa, \tau_y = 2.6\ Pa, k = 2.6\ Pa.s^n$) [24, 37].

## Appendix C: Effect of the location of the outflow boundary

As explained already, when it appears, the transition zone is constantly developing, as the stress magnitude there only asymptotically approaches the yield stress from above. This means that imposing a no-gradient condition normal to the outflow boundary will somewhat affect the solution near this outflow boundary. This boundary condition is typically applied irrespective of the method chosen to numerically solve this problem. We have also tested the open boundary condition [61] at the outflow using the Finite element method and obtained the same results. Hence inevitably, the length and shape of the transition zone will be affected by the location of the outflow boundary. This effect will be examined in this subsection.

(a)                                                                                          (b)

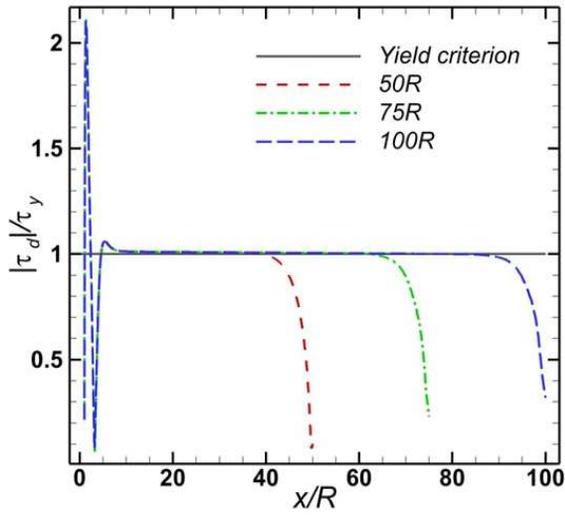
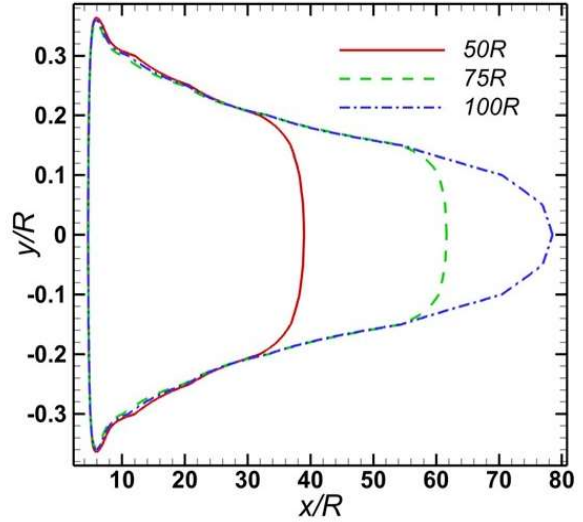

(c)                                                (d)

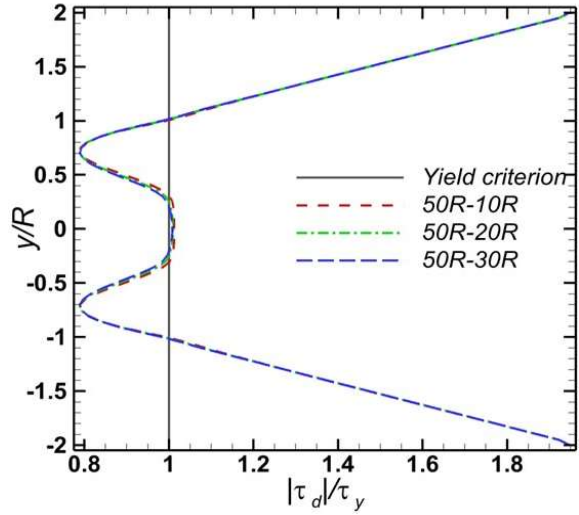
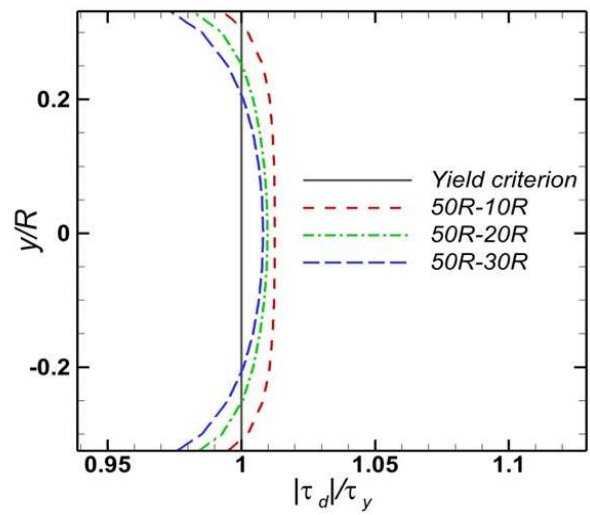

Figure 26. Effect of the location of the outflow boundary on (a) the stress magnitude along the plane of symmetry, (b) the boundary of the transition zone, (c) the variation of the stress magnitude along the $y$-axis, and (d) the same as in (c) focusing near the axis. In (d) the intersection of the vertical solid line with the curves determines the range of the yielded material in the transition zone. The legends in (a) and (b) show the $x$-location of the downstream boundary ($x = 50R, \, 75R,$ and $100R$) and in (c) and (d) they show the $x$-location of the downstream boundary ($x = 50R$) and the $x$-location where the stress is calculated ($x = 10R, 20R,$ and $30R$).

Fig. 26 (a) illustrates the stress magnitude normalized with the yield stress, when the distance of the outflow boundary from the cylinder center increases from $50R$ to $75R$ and finally to $100R$. After a strong variation near the cylinder and the "negative wake" (see also Fig. 6) this ratio slowly approaches unity from above, indicating that we are in the transition zone. For some distance along the $x$-axis, it remains ever so slightly above unity, but eventually reaches it and then abruptly falls below it, indicating that the material becomes unyielded. This ratio becomes almost zero at the outflow boundary, because of the boundary condition applied there. The axial distance at which this ratio intersects unity increases nearly proportionately to the respective distance of the outflow boundary. In any case, Fig. 26 (a) verifies the decrease of the stress magnitude towards the yield stress. Clearly, the location of the exit boundary affects the stress magnitude, and, hence, the material state for the same distance of approximately 10 to 15 radii upstream. Thus, this part of the solution should be ignored, and this effect should be kept in mind in the previous results.

Fig. 26 (b) presents the shapes of the transition zones for the same three locations of the outflow boundary. Their thicknesses decrease almost linearly in the axial direction and their side boundaries remain approximately the same up until each transition zone is about to close. This indicates that in the region that they have the same thickness, the stress components converge to the same value as well. Consequently, in the same region the solution is not significantly affected by the location of the outflow boundary. The corresponding lengths of the transition zones for which their thickness stops decreasing linearly for exit boundary at 50R, 75R, and 100R are $32R$, $55R$, and $71R$, respectively, while at the axis of symmetry they end at $38.9R$, $61.7R$, and $78.5R$.

Fig. 26 (c) and 26 (d) depict the variation of the stress magnitude in the $y$-direction at three axial locations, $10R$, $20R$ and $30R$ for the same length of the outflow boundary, $50R$. While Fig. 26 (c) does not exhibit noticeable differences, focusing on the central region in Fig. 26 (d) reveals that, as we move downstream, the magnitude of the stress decreases ever so slowly approaching the yield value, and decreases the thickness of the yielded area, which becomes approximately $0.62R$, $0.5R$, and $0.4R$, respectively.

**Appendix D:** Mesh convergence study when the predicted flow remains transient or becomes erratic.

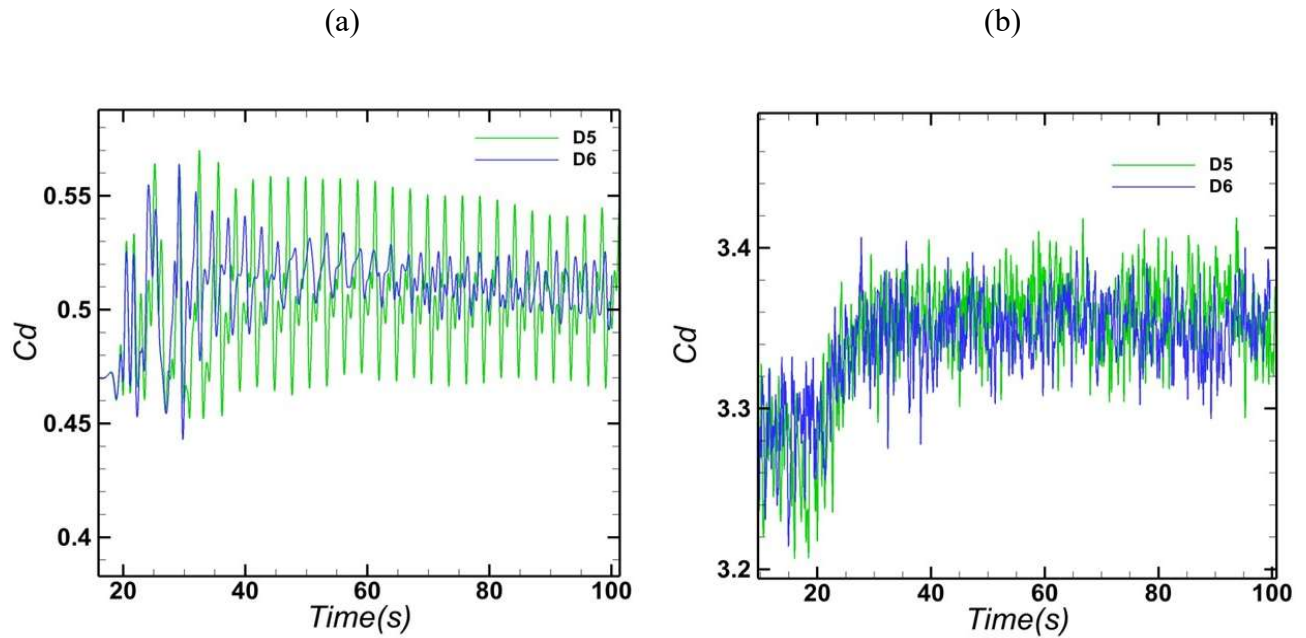

Figure 27. Evolution of the drag coefficient for a) $G = 2\ Pa$ and b) $\tau_y = 37\ Pa$. Some details of each mesh are reported in Table 4.

Fig. 27 illustrates the evolution of the drag coefficient for the highest $Wi$ and $Bn$ cases, which are the most demanding ones using the two meshes (D5 and D6) generated for the transient simulations. For $\tau_y = 37\ Pa$, the oscillation of the drag coefficient shows acceptable convergence with mesh and in time between the two meshes (see Fig. 27 b). However, for $G = 2\ Pa$, the oscillation exhibits different amplitudes, which can be attributed to the abrupt changes in normal elastic stresses near the cylinder. Since increasing $Wi$ directly affects normal elastic forces, the drag coefficient becomes highly sensitive to the number of cells near the cylinder (see Fig. 27 a). For this transient flow, even small variations in mesh resolution in this critical area can significantly impact the results.

We conducted more tests by further refining the mesh or changing locally its refinement, yet differences persisted, particularly in the normal stress components ($\tau_{xx}\ and\ \tau_{yy}$) close to the cylinder. Similar challenges in achieving mesh convergence in viscoelastic fluid flows have been reported in other studies [58, 59]. Based on these findings, we conclude that changes in mesh resolution around the cylinder surface substantially influence the results due to significant stress variations in this region. These observations are consistent with results from mesh convergence tests in the 4 to 1 contraction flow [32], where the lip vortex was similarly sensitive to changes in mesh resolution.

## Appendix E: Data and code availability

Data will be made available upon request. Here is the link to the simulation code:

(https://www.dropbox.com/scl/fi/8ylskcmj0cgl3yop5s7x4/Fluid2.rar?rlkey=twmanzqrohae9h9nanfkbfskn&st=7b97guk6&dl=0).

## CRediT authorship contribution statement

Milad Mousavi: Data curation, Investigation, Validation, Writing original draft, Software. Yannis Dimakopoulos: Conceptualization, Investigation, Methodology, Supervision, Funding acquisition, Writing reviewing & editing. John Tsamopoulos: Conceptualization, Investigation, Methodology, Validation, Supervision, Funding acquisition, Writing, reviewing & editing.

## Acknowledgements

This research has received funding from the European Union´s Horizon 2020 research and innovation programme under the Marie Skłodowska-Curie grant agreement No 955605. The authors gratefully acknowledge the several helpful comments by both reviewers that have led to a better presentation of this work.

## Declaration of interests

The authors report no conflict of interest.